\documentclass[pra,twocolumn,superscriptaddress,longbibliography]{revtex4}
\usepackage[latin9]{inputenc}
\usepackage{color}
\usepackage{amsmath}
\usepackage{amssymb}
\usepackage{bm}
\usepackage{braket}
\usepackage{graphicx}
\usepackage{tikz}
\usepackage{mathrsfs} 
\usepackage{float}
\usepackage{bbm}
\usepackage{epstopdf}
\usepackage{epsfig}
\usepackage{verbatim}
\usepackage{array}
\usepackage{setspace}
\usepackage{times}
\usepackage{algorithm}
\usepackage{algpseudocode}
\usepackage{multirow} 
\usepackage{booktabs} 
\usepackage{siunitx} 

\usepackage[unicode=true,
 bookmarks=true,bookmarksopen=false,
 breaklinks=false,pdfborder={0 0 1},backref=false,colorlinks=true]
 {hyperref}
\hypersetup{
 linkcolor=magenta, urlcolor=blue, citecolor=blue, pdfstartview={FitH}, hyperfootnotes=false, unicode=true}

\newcommand\blue[1]{{\color{black}#1}}

\newcolumntype{C}[1]{>{\centering\arraybackslash$}p{#1}<{$}}

\begin{document}

\title{Hamiltonian Learning via Inverse Physics-Informed Neural Networks}

\author{Jie Liu}
\affiliation{Department of Physics, City University of Hong Kong, Tat Chee Avenue, Kowloon, Hong Kong SAR, China}
\affiliation{City University of Hong Kong Shenzhen Research Institute, Shenzhen, Guangdong 518057, China}
\affiliation{Quantum Science Center of Guangdong-Hong Kong-Macao Greater Bay Area, Shenzhen, Guangdong 518045, China}
\author{Xin Wang}
\email{x.wang@cityu.edu.hk}
\affiliation{Department of Physics, City University of Hong Kong, Tat Chee Avenue, Kowloon, Hong Kong SAR, China}
\affiliation{City University of Hong Kong Shenzhen Research Institute, Shenzhen, Guangdong 518057, China}

\date{\today}

\date{\today}

\begin{abstract}

Hamiltonian learning (HL), enabling precise estimation of system parameters and underlying dynamics, plays a critical role in characterizing quantum systems. However, conventional HL methods face challenges in noise robustness and resource efficiency, especially under limited measurements. In this work, we present \textit{Inverse Physics-Informed Neural Networks for Hamiltonian Learning (iPINN-HL)}, an approach that \blue{incorporates the Schr\"{o}dinger equation as a soft constraint via a loss function penalty into the ML procedure.} This formulation allows the model to integrate both observational data and known physical laws to infer Hamiltonian parameters with greater accuracy and resource efficiency. We benchmark iPINN-HL against a deep-neural-network-based quantum state tomography method (denoted as DNN-HL) and demonstrate its effectiveness across several different scenarios, including one-dimensional spin chains, cross-resonance gate calibration, crosstalk identification, and real-time compensation to parameter drift. Our results show that iPINN-HL can approach the Heisenberg limit and exhibits robustness to noises, while outperforming DNN-HL in accuracy and resource efficiency. 
Therefore, iPINN-HL is a powerful and flexible framework for quantum system characterization for practical tasks.

\end{abstract}

\maketitle

\section{INTRODUCTION} \label{Intro}

 Over the past few decades, advances in quantum technologies, particularly in the field of quantum control, have enabled substantial progress in precise manipulation of quantum states~\cite{dridi2020optimal,chen2024universal,ball2021software,sivak2022model,mohan2023robust}. Nonetheless, achieving even higher precision remains crucial for the design of quantum gates, refinement of control strategies, and robust implementation of high-fidelity quantum operations under realistic, noisy conditions. Such improvements are essential for realizing practical quantum error correction and for scaling up viable quantum computing architectures~\cite{acharya2024quantum,bravyi2024high}. Reaching this goal demands more advanced system characterization techniques capable of accurately identifying device parameters and diagnosing noise sources~\cite{chen2024benchmarking,kanazawa2023qiskit,pasquale2023towards,wyderka2023complete,dahlhauser2024benchmarking}.

Current quantum devices are subject to various sources of noise, including decoherence, readout errors, and unintended interactions between neighboring qubits~\cite{khan2024multiaxis,urbanek2021mitigating,zhao2022quantum}. Identifying and quantifying these error sources is a key step toward effective device characterization. Moreover, quantum hardware often requires frequent recalibration to mitigate parameter drift over time~\cite{proctor2020detecting}, which poses significant challenges to the efficiency and stability of system characterization procedures.

System characterization and Hamiltonian learning (HL) are intrinsically related, as both aim to uncover the underlying dynamics of quantum systems~\cite{wang2017experimental}. In HL, the objective is to infer the Hamiltonian that governs the evolution of a system from experimental data, using limited classical and quantum resources~\cite{an2024unified,yu2023robust,gu2024practical,dutkiewicz2024advantage}. In quantum mechanics, the evolution is governed by the Schr\"{o}dinger equation, in which both the Hamiltonian and various noise terms determine the dynamics of the system. Consequently, quantum system characterization can be formulated as a HL problem, where the goal is to extract an accurate description of the dynamics of the system from data. Learning the Hamiltonian provides access to system parameters, interactions, and noise sources, thereby laying the groundwork for effective calibration and precise control.

However, it is unavoidable that the resources required to learn a generic many-body Hamiltonian grow exponentially with the system size \cite{mohseni2008quantum}. When the system size is fixed, the estimation accuracy scales polynomially with the resources as $m^{-\ell}$ and is bounded by the standard quantum limit with $\ell=1$ or  by the Heisenberg limit with $\ell=2$, where $m$ quantifies the amount of resources used in estimation.
While it is possible for an optimized HL protocol to surpass the standard quantum limit, practical limitations such as imperfect state preparation, manipulation, and various sources of noise in real devices often hinder the full attainment of the Heisenberg limit. Despite these challenges, it remains advantageous to minimize the resources needed to achieve a given level of accuracy within the HL framework. This focus on resource efficiency is a central aim of this paper.

Machine Learning (ML) has been widely applied in HL. In HL, the requirement to gather extensive measurement data over time to accurately characterize system dynamics provides an ideal context for ML. As a data-driven approach, ML is particularly effective at identifying patterns and inferring hidden parameters from large datasets, even when faced with noise or incomplete information.
Recent studies have applied various ML techniques to extract system Hamiltonians both numerically and experimentally, including recurrent neural networks \cite{flurin2020using}, autoencoders \cite{tucker2024hamiltonian}, Bayesian inference \cite{evans2019scalable}, and active learning \cite{dutt2023active}. For example, neural networks can be trained in real-time to predict the quantum trajectory of a superconducting qubit undergoing unitary evolution \cite{flurin2020using}. While traditional data-driven ML methods have demonstrated considerable success in HL protocols, they often lack direct integration of the physical principles that govern the system. \blue{This limitation significantly increases the data demand for achieving reliable generalization and hinders the adaptability of the model when extrapolating beyond the observed data regime.} These issues are particularly pronounced in HL, where data collection can be costly, and densely sampling over the entire time period is impractical. 

Recently, Physics-Informed Neural Networks (PINNs), an interesting approach in ML that goes beyond purely data-driven methods, address this gap by incorporating the underlying physical equations into the learning process \cite{raissi2019physics,cai2021physics, cuomo2022scientific, Castelano.24, Davoodi.25}.
\blue{By integrating \textit{a priori} physical knowledge---}such as the Schr\"{o}dinger equation\blue{---into ML, PINNs accelerate convergence, enhance accuracy, and enable more reliable extrapolation beyond observed data \cite{zhu2023reliable,yang2019predictive}.}
Ref.~\cite{zhou2024data} applies PINN to infer material properties in heat conduction and to reconstruct flow fields in fluid dynamics. PINN is also used in Ref.~\cite{chen2020physics} to reconstruct permittivity and shape of nanostructures from observed scattering data.  This approach is particularly advantageous in scenarios where data is sparse or noisy, as the embedded physical principles help regularize the model and guide it toward physically meaningful solutions. 

To utilize PINN in HL, we propose an algorithm called  inverse Physics-Informed Neural Network for HL (iPINN-HL). In iPINN-HL, the problem of HL is first reformulated as the inverse problem of Partial or Ordinary Differential Equation (PDE/ODE). 
Unlike the forward problem of PDE/ODEs, where we solve the equations with known parameters and boundary conditions to numerically find the solution at various time points, 
the goal of the inverse problem is to estimate the unknown parameters in the PDE/ODE such that the numerical solution obtained under these estimated parameters aligns with experimental observations. \blue{HL can be naturally formulated as an inverse problem, governed by the Schr\"odinger equation, with the Hamiltonian and noise terms treated as unknown parameters.}

\blue{In this work, we numerically evaluate the performance of iPINN-HL across a range of tasks. We find that iPINN-HL can approach the Heisenberg limit under ideal conditions, i.e., with unconstrained state preparation and measurement and purely unitary dynamics. We further evaluate its performance in practical calibration tasks, including cross-resonance gate calibration, crosstalk quantification, and parameter-drift compensation. Our results show that it consistently outperforms the purely data-driven baseline, DNN-HL, which estimates the unknown parameters using a deep neural network (DNN).}

\blue{Our iPINN-HL framework introduces soft physics-informed constraints through the loss function, providing flexibility in handling noisy and sparse quantum data. In comparison, structure-preserving networks such as SympNets \cite{jin2020sympnets} enforce Hamiltonian dynamics exactly by preserving symplectic structure, ensuring energy conservation but at the cost of architectural complexity. Hamiltonian Neural Networks \cite{greydanus2019hamiltonian} offer another alternative by learning dynamics via time-derivative prediction, balancing generality with accuracy. Unlike these strict formulations, iPINN-HL favors robustness and adaptability over exact Hamiltonian preservation, making it particularly suitable for quantum systems where data acquisition is costly, as demonstrated in Sec.~\ref{sec:results}.}


The paper is organized as follows: In Sec.~\ref{sec:HL}, we provide an overview of HL, introducing the theoretical framework of DNN-HL and the concept of inverse problems. We then discuss the principles of PINNs and their relevance to HL tasks. In Sec.~\ref{sec:PINN}, we present iPINN-HL in detail, explaining its formulation, the integration of physics loss, and the neural network architecture employed. Sec.~\ref{sec:results} shows the application of iPINN-HL across various scenarios, including one-dimensional spin chain systems, cross-resonance gate calibration, crosstalk identification, and parameter drift compensation, with comparative analysis against DNN-HL. Finally, in Sec.~\ref{sec:conclusion}, we summarize the findings, highlight the advantages of iPINN-HL, and discuss potential further developments.


\section{Preliminaries}\label{sec:preliminaries}
\subsection{Hamiltonian Learning and Maximum Likelihood Estimation}\label{sec:HL}
\begin{figure}
	\centering
    \includegraphics[width=\linewidth]{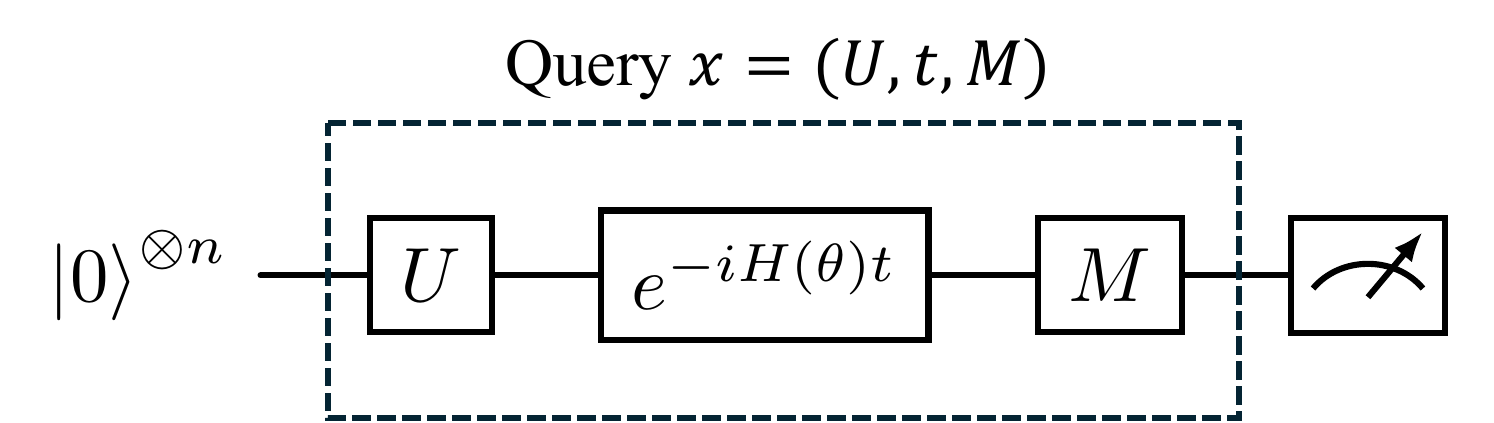}
    \caption{Circuit illustration of the quantum query model. After the a query $x=(U,t,M)$ is input to the system, the output is an $n$-bit string.}
    \label{fig:query_model}
\end{figure}

\begin{figure*}[t]
    \centering
    \includegraphics[width=\textwidth]{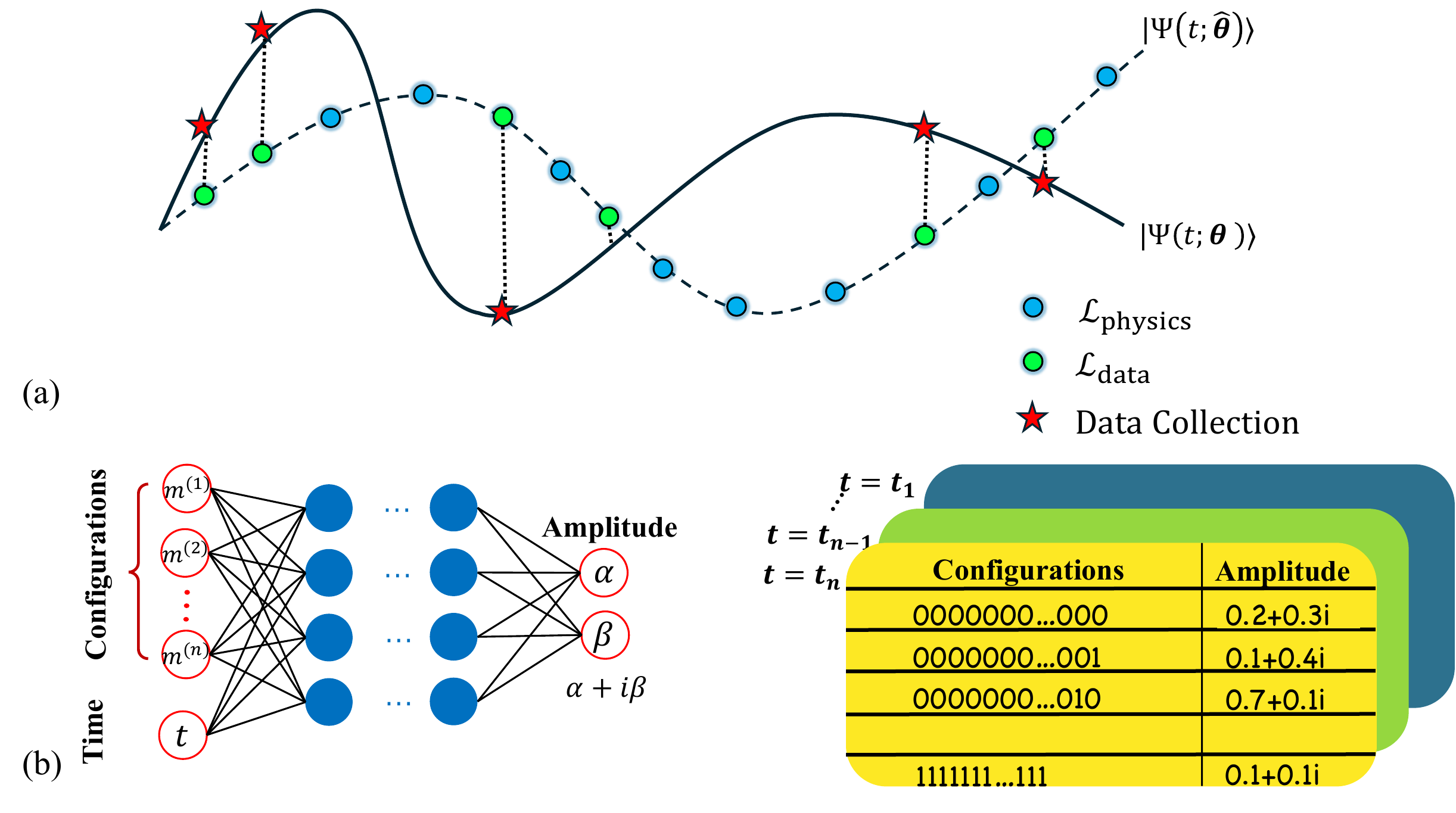}
    \caption{(a) Visualization of Physics-Informed Neural Networks (PINNs) for HL. The solid curve shows the true solution $\ket{\Psi(t; \bm{\theta})}$, while the dashed curve represents the estimated solution $\ket{\Psi(t; \bm{\hat{\theta}})}$. Red stars mark data points, green dots indicate prediction errors, and blue dots enforce physical laws by ensuring consistency with the Schr\"{o}dinger equation. (b) Neural Network Quantum States (NNQS) representation and its tabular form at different time points. \blue{The output nodes $\alpha$ and $\beta$ of the neural network represent the real and imaginary parts, respectively, of the complex amplitude of the quantum state $\braket{m| \Psi(t)}$ at time $t$.} The neural network efficiently captures the complex amplitudes of quantum states across various configurations, demonstrating high expressive capacity for representing many-body quantum systems. The table on the right shows the amplitudes of different configurations at discrete time steps $ t_1, t_2, \dots, t_n $. Automatic differentiation of NNQS enables efficient computation of time derivatives of the quantum state, facilitating the enforcement of dynamical constraints of Schr\"{o}dinger equation.
}
\label{fig:lossandNNQS}
\end{figure*}

Let $ H $ be the Hamiltonian of an $ n $-qubit system. It can be expressed as a linear combination of $ n $-qubit Pauli strings:
\begin{equation}
    H = \sum_{j=1}^{4^n} \theta_j P_j,
\end{equation}
where $ P_j \in \{I, \sigma_x, \sigma_y, \sigma_z\}^{\otimes n} $ and $ \theta_j = \frac{1}{2^n} \mathrm{Tr}(H P_j) $. Here, $ \sigma_x $, $ \sigma_y $, and $ \sigma_z $ are the Pauli matrices, and $I$ is the identity matrix. In HL, the problem to learn the Hamiltonian of the system can be reduced to estimating a set of unknown parameters $\bm{\theta}=\{\theta_1,\theta_2,...,\theta_{4^n}\}$ from experimental data. Data collection procedure of HL can be formulated as a query model defined in Ref.~\cite{dutt2023active}. A query $x$ consists of three components $U$, $t$ and $M$, as illustrated in Fig.~\ref{fig:query_model}: 
\begin{itemize}
	\item \textit{The state preparation unitary $U$.}  The initial state $\ket{\psi_0}$ is specified by $\ket{\psi_0}=U\ket{0}^{\otimes n}$. In principle, $\ket{\psi_0}$ can be any unit vector in the Hilbert space, and this is generally assumed in most HL protocols. However, in the context of real device calibration, greater attention must be paid to the constraints on the unitary $U$. For instance, in two-qubit gate calibration, it is not permissible to assume that the initial state $\ket{\psi_0}$ is entangled. In this case, the allowed  $U$ must have a tensor product form, i.e., $U=U_1 \otimes U_2$. Additionally, since the execution time of single-qubit gates is much shorter than the typical timescale of HL protocols, and the calibration of such gates is generally more accurate, it is often reasonable to assume that the state preparation process is not significantly affected by noise sources. Thus, we can consider the state preparation unitary $U$ as ideal for practical purposes. For a practical example of device calibration, see Ref.~\cite{wittler2021integrated}.
	\item \textit{The controlling parameter $t$.} 
	The controlling parameter $t$ refers to the time Hamiltonian $H$ being applied to the system. In a typical HL protocol, $t$ should take different values so that the dynamics during the time span can be accurately characterized. Algorithms exist to choose $t$ systematically to reduce resources needed in HL \cite{dutt2023active}. However, in this work, we keep the controlling parameters equally spaced by $\Delta t$ to examine the extrapolation ability of iPINN-HL. In other words, we aim to examine whether the embedded physics knowledge can increase the ability to infer unobserved data at other controlling parameter $t$, and in turn, reduce the resources required in HL.
	\item \textit{The measurement operator $M$.} The measurement operator specifies the set of projective measurement bases $\{\ket{\phi_m}\}$. In this work, we assume that the final measurement bases are always fixed as the computational bases $\{\ket{m}\}$, therefore, $M=\sum_m \ket{\phi_m}\bra{m}$. Similar to the constraint on the state preparation unitary $U$ in two-qubit gate calibration, $\ket{\phi_m}$ should not be an entangled state.
\end{itemize}

\blue{
We note that our definition of the measurement operator, $M = \sum_m \ket{\phi_m}\bra{m}$, differs from the standard formalism, where projectors take the form $\ket{\phi_m}\bra{\phi_m}$.  In our construction, the computational basis $\ket{m}$ is fixed as the measurement basis, so that the outputs are always bit-strings. This choice facilitates data processing in the neural network for iPINN-HL's loss term.
}

\begin{algorithm}[b]
\caption{Deep Neural Network-based Hamiltonian Learning (DNN-HL)}
\label{alg:dnn_hl}
\begin{algorithmic}[1]

\Require Dataset $D$, initial state $\ket{\psi_0}$, learning rate $\eta$, total iterations $J$, and ADAM optimizer
\Ensure Estimated Hamiltonian parameters $\bm{\hat{\theta}}$

\State Initialize neural network weights $w$ and Hamiltonian parameters $\bm{\hat{\theta}}$

\For{$\mathrm{Epoch} = 1$ to $J$}
	\State Group dataset $\tilde{D} = \{(U_k, t_k, M_k, \{y_k^{(s)}\}_{s=1}^{S_k})\}_{k=1}^N$, where each group shares the same control parameters but includes multiple measurement outcomes;
    \For{Each measurement configuration $(U_k, t_k, M_k, \{y_k^{(s)}\})$ in $\tilde{D}$}
        \State Use the DNN to reconstruct quantum state $\ket{\phi_k}$ from the measurement outcomes $\{y_k^{(s)}\}_{s=1}^{S_k}$
        \State Compute the model-evolved state: $\ket{\psi_k(\bm{\hat{\theta}})} = e^{-iH(\bm{\hat{\theta}})t_k}U_k \ket{\psi_0}$
        \State Compute data loss: $\mathcal{L}_\mathrm{data}^{(k)} = \left\| \ket{\phi_k} - \ket{\psi_k(\bm{\hat{\theta}})} \right\|^2$
    \EndFor
    \State Combine total loss: $\mathcal{L}_\mathrm{total} = \sum_k \mathcal{L}_\mathrm{data}^{(k)}$
    \State Update $w$ and $\bm{\theta}$ using ADAM optimizer with learning rate $\eta$
\EndFor

\State \Return Optimized Hamiltonian parameters $\bm{\hat{\theta}}$
\end{algorithmic}
\end{algorithm}

The system to be characterized receives a query 
\begin{equation}
x=(U,t,M),\label{eq:query}
\end{equation}
and output a single shot measurement result $y$, which is an $n$-bit string based on conditional probability defined as:

\begin{equation}
    p(y|x;\bm{\theta})=\bigg|\bra{y}Me^{-iH\blue{(\bm{\theta})}t}U\ket{0}^{\otimes n}\bigg|^2.
\end{equation}

After collecting a dataset $D=\{(x_j,y_j)\}_{j=1}^{N}$, where $j$ \blue{indexes each entry}, we aim to estimate the unknown parameters $\bm{\theta}$. \blue{A natural approach is Maximum Likelihood Estimation (MLE), which produces an estimate} $\bm{\hat{\theta}}$ given by:
\begin{equation}
    \bm{\hat{\theta}}=\underset{\bm{\theta}}{\operatorname{arg\,max}} \frac{1}{N} \sum_k^{N} \log{p(y_k|x_k;\bm{\theta})}.
\end{equation}
\blue{Accordingly, we define} the loss function $\mathcal{L}(\bm{\theta}|D)$  as:
\begin{equation} 
	\mathcal{L}(\bm{\theta}|D)=-\frac{1}{N} \sum_k^{N}\log{p(y_k|x_k;\bm{\theta})}, \label{eq:dataloss_DNN-HL}
\end{equation}
and by minimizing the loss function, we find the estimated parameter $\bm{\hat{\theta}}$. 

The DNN-HL method is adapted from the neural network-based quantum state tomography method proposed in Ref.~\cite{torlai2018neural}, where a Restricted Boltzmann Machine (RBM) is used to reconstruct the quantum state from measurement data. In our work, we extend this idea to HL  by learning the underlying Hamiltonian parameters from the reconstructed state. To align the model class with that of iPINN-HL and enable a fair comparison, we replace the RBM with a deep neural network. Although the architectures differ, both RBMs and DNNs are capable of representing complex distributions and learning nontrivial mappings from data under sufficient capacity. This makes the replacement legitimate in the context of benchmarking model performance for HL tasks.
After the DNN is applied to reconstruct a set of state $\{\ket{\phi_i}\}$ for every controlling time parameter $t_i$ by following the same routine demonstrated in Ref.~\cite{torlai2018neural}, we minimize the discrepancy between the reconstructed state $\ket{\phi_i}$ and the estimated state $e^{-iH(\bm{\hat{\theta}})t_i}\ket{\psi_0}$ at time parameter $t_i$. Hence the loss function  can be explicitly \blue{written} as:
\begin{equation}
	\mathcal{L}_\mathrm{data}^\mathrm{DNN}(\blue{\bm{\hat{\theta}}}|D)=\sum_i \bigg|\ket{\phi_i}-e^{-iH(\bm{\hat{\theta}})t_i}\ket{\psi_0} \bigg|^2, \label{eq:loss_dnn}
\end{equation}
for all the initial state $\ket{\psi_0}$ generated by the state preparation unitary $U$.

 The details of the DNN-HL algorithm are provided in Algorithm~\ref{alg:dnn_hl}.

\subsection{Inverse Problem and PINNs}\label{sec:PINN}
HL can be viewed as an inverse problem of a PDE--specifically, the Schr\"odinger equation:
\begin{equation}
	i\hbar\frac{d\ket{\Psi(t)}}{dt}=H(\bm{\theta})\ket{\Psi(t)}, \label{eq:schrodinger}
\end{equation}
where the goal is to infer unknown Hamiltonian parameters from observed dynamics and measurements. For simplicity, we assume the state is pure in Eq.~\eqref{eq:schrodinger}, therefore the dynamics is unitary. This assumption applies to the remainder of this chapter as well as  Sec.~\ref{sec:spinchain}, Sec.~\ref{sec:Crosstalk} and Sec.~\ref{sec:drift}. More complicated cases involving noises and therefore mixed states, are considered in in Sec.~\ref{sec:CR_gate}.


Fig.~\ref{fig:lossandNNQS}(a) illustrates the fundamental concept of PINNs, supplemented with the Schr\"odinger equation.
The solid curve represents the true solution of the system $\ket{\Psi(t;\bm{\theta})}$, where $\bm{\theta}$ denotes system parameters. The dashed curve represents the estimated solution of the system $\ket{\Psi(t;\bm{\hat{\theta}})}$ obtained under the proposed parameter estimation $\bm{\hat{\theta}}$. The red stars indicate points where actual observed data is collected, and the model minimizes the discrepancy (shown by the vertical z lines) between predictions (green dots) and real-world measurements. The blue dots represent locations where the physical laws are enforced. At these points, the first derivative of the estimated state, $\frac{\partial \ket{\Psi(t;\bm{\hat{\theta}})}}{\partial t}$, is computed alongside $\ket{\Psi(t;\bm{\hat{\theta}})}$ to ensure that the system satisfies the Schr\"{o}dinger equation in Eq.~\eqref{eq:schrodinger}. Importantly, no data from the true solution is required at these points, as the physical laws themselves guide the learning process.

In order to utilize PINN to solve the HL problem, we first need to represent the quantum state $\ket{\Psi(t)}$ using a neural network. Accurately representing a quantum state can be computationally challenging due to the exponential scaling of the state space with the number of particles $n$ involved, illustrated in the right panel of Fig.~\ref{fig:lossandNNQS}(b), where one table corresponds to one time point and every table consists of $2^n$ entries. 
We address this challenge by parameterizing the quantum states with neural networks, denoted as Neural Network Quantum States (NNQSs). This method facilitates efficient approximations of states that would otherwise demand extensive computational resources \cite{jia2019quantum,gutierrez2022real}. In NNQS, the neural network acts as a parameterized wave function $\Psi_{w}(t,m)$ where $w$ is the trainable weight in the network. The input is time $t$ and the configuration $m$ which is the bit-string representation of the chosen computational basis $\ket{m}$, and its outputs correspond to the amplitudes of $\ket{\Psi(t)}$ in the chosen basis $\ket{m}$, $\braket{m | \Psi(t)}$. Therefore, NNQS allows for a continuous and differentiable approximation of the state across time. With the help of back-propagation, the NNQS can easily calculate and get the first order differentiation of $\ket{\Psi(t)}$ with respect to time $t$, i.e., $\frac{d\ket{\Psi(t)}}{dt}$. 

Implementing PINN, we randomly sample $P$ points (which we call the number of \blue{\emph{collocation points}}  \cite{bauduin2025impact} hereafter) in the time domain $(t_1,t_2,...,t_j,...,t_P)$. We incorporate new loss function (``physics loss'') in the training process (termed as $\mathcal{L}_\mathrm{physics}$):
\begin{equation}
	\mathcal{L}_\mathrm{physics}(w,\bm{\theta})=\blue{\frac{1}{P}}\sum_j^P \bigg|i\hbar\frac{d\ket{\Psi(t_j)}}{dt}-H(\bm{\theta})\ket{\Psi(t_j)}\bigg |^2. \label{eq:PINN_phy}
\end{equation}
\blue{The full wave function $\ket{\Psi(t_j)}$ is assembled from the $2^n$ outputs of neural network,
\begin{equation}
	\ket{\Psi(t_j)} = \sum_m (\alpha_m+i\beta_m)\ket{m},
\end{equation}
where $\alpha_m$ and $\beta_m$ are the two outputs of neural network given input being configuration $\ket{m}$ and time $t_j$.
}

For every different initial quantum state $\ket{\Psi^{(l)}(0)}$, we use one NNQS $\Psi_{w}^{(l)}(t,m)$. The NNQS must satisfy the initial condition of Eq.~\eqref{eq:schrodinger}. Hence the loss function about initial condition termed as $\mathcal{L}_\mathrm{initial}$ is also introduced in the training, defined as follows:
\begin{equation}
	\mathcal{L}_\mathrm{initial}(w^{(l)},\bm{\theta}|D)=\frac{1}{K} \sum_m \bigg|\braket{m|\Psi^{(l)}(0)}-\Psi_{w}^{(l)}(t=0,m) \bigg|^2, \label{eq:PINN_initial}
\end{equation} 
where $K$ is the total number of terms in the summation and $w^{(l)}$ is the weight of $l$-th NNQS.

Together with a data loss function $\mathcal{L}_\mathrm{data}$  defined similarly to Eq.~\eqref{eq:dataloss_DNN-HL} :
\begin{equation}
	\mathcal{L}_\mathrm{data}(w,\bm{\theta}|D)=-\frac{1}{N} \sum_k^{N} \bigg| \Psi_w(t_k,y_k) \bigg|^2, \label{eq:PINN_data}
\end{equation}
where $y_k$ is the observed bit-string in the measurement result and $t_k$ is the controlling parameter stored in the $k$-th entry $(x_k,y_k)$ of dataset $D$, we have the total loss function used in training PINN for HL:
\begin{equation}
\begin{aligned}
	\mathcal{L}_\mathrm{total}(w,\bm{\theta}|D)&=\mathcal{L}_\mathrm{data}(w,\bm{\theta}|D)+\lambda_1\mathcal{L}_\mathrm{physics}(w,\bm{\theta})\\
	&+\lambda_2\mathcal{L}_\mathrm{initial}(w,\bm{\theta}|D), \label{eq:PINN_total}
\end{aligned}
\end{equation}
where $\lambda_1$ and $\lambda_2$ are weighting factors. Note that $\mathcal{L}_\mathrm{physics}(w,\bm{\theta})$ is not dependent on dataset $D$. Rather, it is incorporated to enforce that the evolution of NNQS satisfies the Schr\"{o}dinger equation. In order to compare the accuracy of both algorithms under various settings, we use mean squared errors (MSE):
\begin{equation}
	\blue{\mathrm{MSE}=\frac{1}{K}\sum_i(\hat{\theta_i}-\theta_i)^2}.
\end{equation}
\blue{Here, $\theta_i$ denotes the true value of the $i^\mathrm{th}$ parameter to be estimated in HL, which may arise from the Hamiltonian or the noise terms. Its estimate is denoted by $\hat{\theta}_i$, and $K$ is the total number of parameters. We also report the relative mean squared error (Relative MSE) as a performance metric:
\begin{equation}
	\mathrm{Relative\ MSE}=\frac{\sum_i(\hat{\theta_i}-\theta_i)^2}{\sum_i \theta_i^2}.
\end{equation}
}

\begin{figure*}[t]
    \centering
    \includegraphics[width=\textwidth]{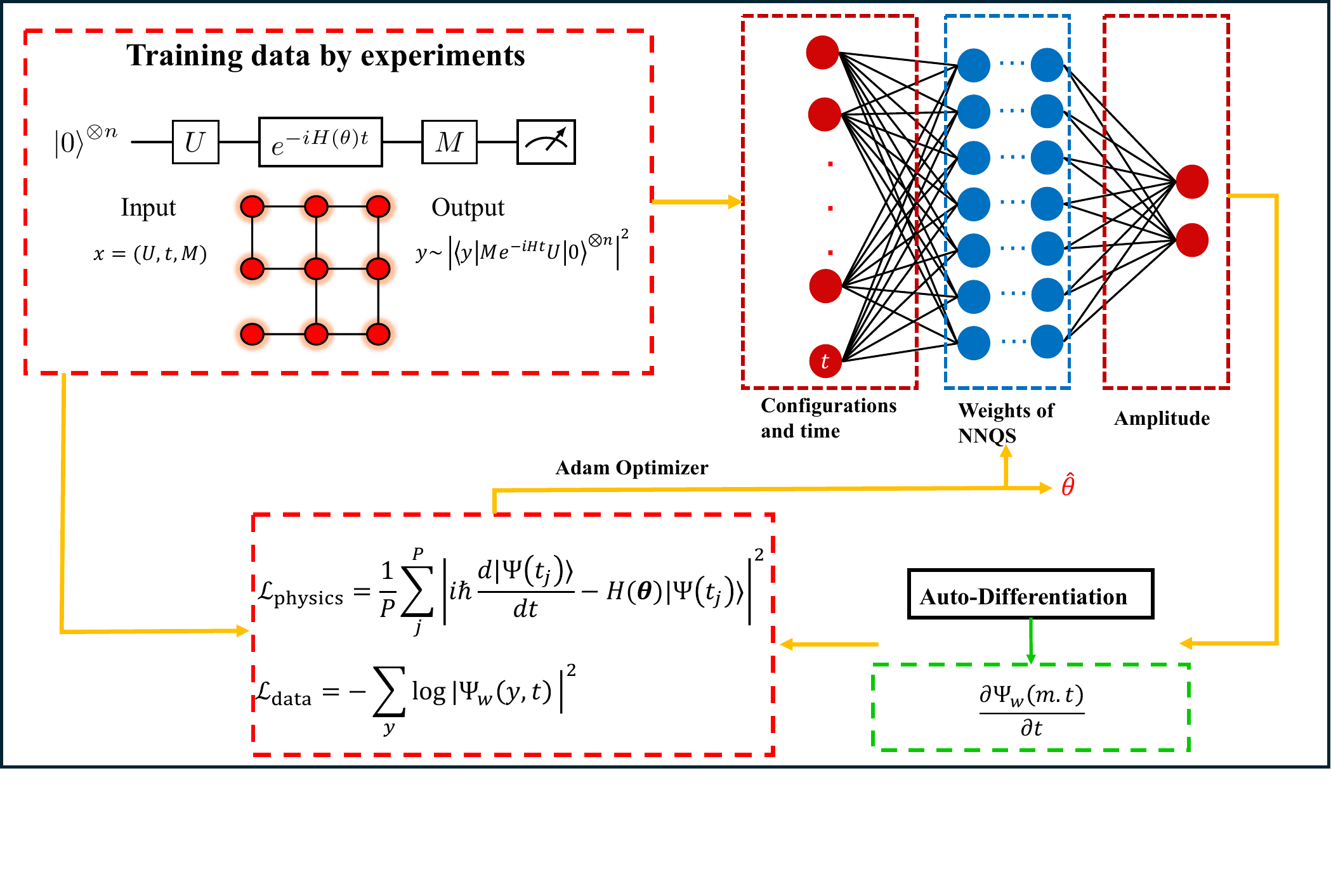}
    \caption{Flowchart illustrating the use of Physics-Informed Neural Networks (PINNs) for HL. Experimental data is generated by evolving an initial quantum state $ \ket{0}^{\otimes n} $ under a unitary transformation $ U $ and time evolution $ e^{-iH(\bm{\theta})t} $, followed by measurement $ M $. The neural network takes configurations and time as inputs and outputs the amplitude of the quantum state, with its weights $ w $ and estimation $\bm{\hat{\theta}}$ optimized using the Adam optimizer. The loss function combines a physical loss $ \mathcal{L}_{\text{physics}} $, which ensures consistency with the Schr\"{o}dinger equation, and a data loss $ \mathcal{L}_{\text{data}} $, which minimizes discrepancies with experimental measurements. Automatic differentiation efficiently computes the time derivative $ \partial \Psi_w(m, t)/\partial t $, facilitating the enforcement of physical constraints during training.}
\label{fig:pipeline}
\end{figure*}

We present  iPINN-HL  in Algorithm \ref{alg:ipinn_hl} and  illustrate \blue{iPINN-HL} in Fig.~\ref{fig:pipeline}.

\begin{algorithm}
\caption{Inverse-Physics Informed Neural Network for Hamiltonian Learning (iPINN-HL)}
\label{alg:ipinn_hl}
\begin{algorithmic}[1]

\Require Dataset $D = \{(x_k, y_k)\}_{k=1}^N$ with $x_k = (U_k, t_k, M_k)$, initial conditions $\ket{\Psi^{(l)}(0)}$, total sampled points $P$, total HL protocol duration $T$, learning rate $\eta$, total iterations $J$ and ADAM optimizer
\Ensure Hamiltonian parameters $\bm{\hat{\theta}}$

\State Initialize parameters $\bm{\hat{\theta}}$ and neural network weights $\{w^{(l)}\}$
\For{$ \mathrm{Epoch} = 1$ to $J$}
    \For{Each initial condition $\ket{\Psi^{(l)}(0)}$}
        \State Sample $P$ time points $\{t_j\}_{j=1}^P$ in time domain $[0,T]$
        \For{each time point $t_j$} 
            \For{\blue{each configuration $m = 1$ to $2^n$} }
                \State Compute NNQS amplitude $\Psi_w^{(l)}(t_j, m)$ and derivative $\frac{d\Psi_w^{(l)}(t_j, m)}{dt}$
             \EndFor
          \State \blue{Assemble the full quantum state $\ket{\Psi(t_j)}=\sum_m \Psi_w^{(l)}(t_j, m)\ket{m}$ and its derivative $\frac{d\ket{\Psi(t_j)}}{dt}=\sum_m d\frac{\Psi_w^{(l)}(t_j, m)}{dt}\ket{m}$}
        \State Calculate \textbf{Physics Loss} according to Eq.~\eqref{eq:PINN_phy}
        \State Calculate \textbf{Initial Condition Loss} according to Eq.~\eqref{eq:PINN_initial}
        \EndFor
    \EndFor
    \State Calculate \textbf{Data Loss} for all data points in $D$ according to Eq.~\eqref{eq:PINN_data}
    \State Combine \textbf{Total Loss} according to Eq.~\eqref{eq:PINN_total}
    \State Update $\bm{\theta}$ and $\{w^{(l)}\}$ using ADAM optimizer
\EndFor
\State \Return Optimized Hamiltonian parameters $\bm{\hat{\theta}}$
\end{algorithmic}
\end{algorithm}

\section{RESULTS} \label{sec:results}
In this section, we present the results of iPINN-HL  to various systems, including 
one-dimensional spin chains (Sec.~\ref{sec:spinchain}), cross-resonance gates  (Sec.~\ref{sec:CR_gate}), crosstalk effects (Sec.~\ref{sec:Crosstalk}), and compensation of parameter drifts  (Sec.~\ref{sec:drift}). 
\subsection{One Dimensional Spin Chains}\label{sec:spinchain}
In this subsection, we apply  iPINN-HL to estimate the Hamiltonian of a one-dimensional $N$-spin chain: 
\begin{equation}
	H_{\mathrm{spin}}=\sum_{\langle i,j\rangle}J_{ij}\sigma_z^{(i)}\sigma_z^{(j)}+\sum_i \omega_i \sigma_z^{(i)},
\end{equation}
where $1\le i,j\le N$ label the spins, $\langle i,j\rangle$ indicates the nearest neighbor. The interaction strength ($J_{ij}$) and  local external field ($\omega_i$) are parameters to be reconstructed. We further implement the periodic boundary condition so that $\langle N,1\rangle$ is a pair of  nearest-neighbors.

\begin{figure*}[t]
    \centering
    \includegraphics[width=\textwidth]{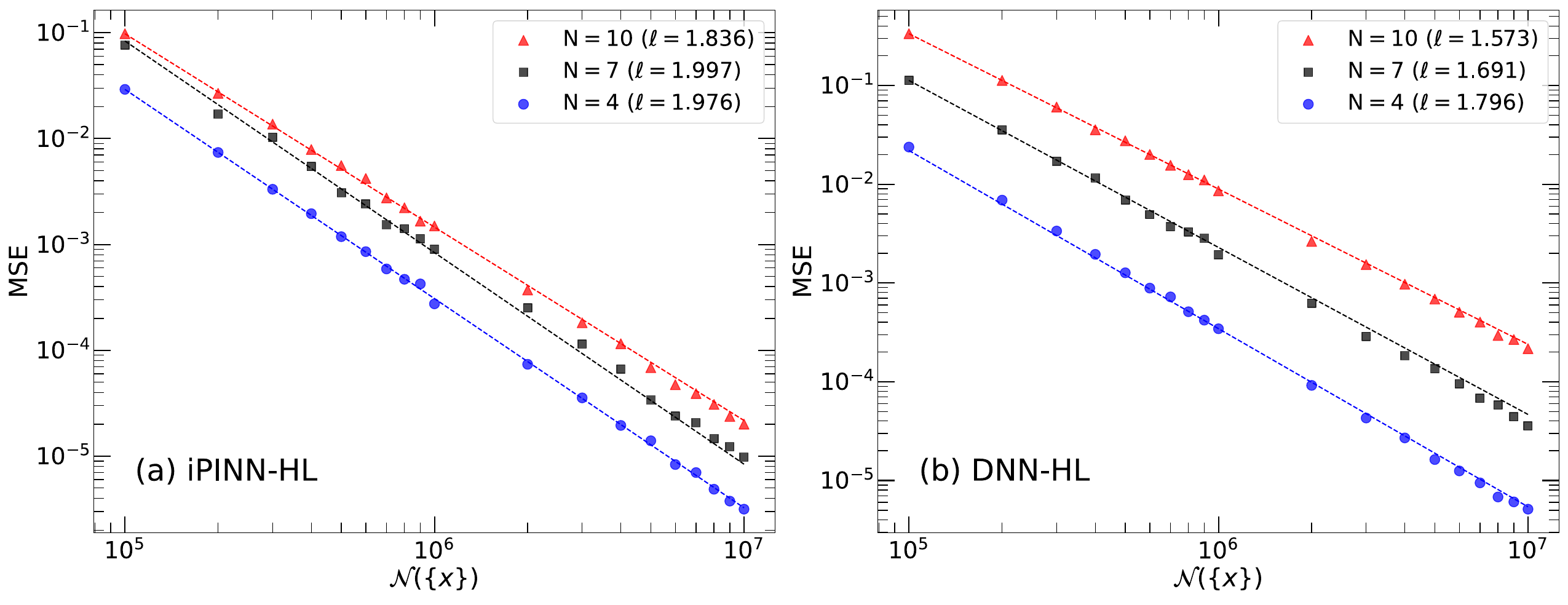}
    \caption{The scaling plot of MSE of iPINN and DNN-HL with respect to the number of entries in the dataset $D$ in (a) and (b) respectively when the number of spins in one-dimensional spin chain is 4, 7 and 10. Here we set $J=1$, $\omega=0.5$ and the number of \blue{collocation points} $P=50$.
    }\label{fig:scaling_N}
\end{figure*}

\begin{figure}
	\centering
    \includegraphics[width=\linewidth]{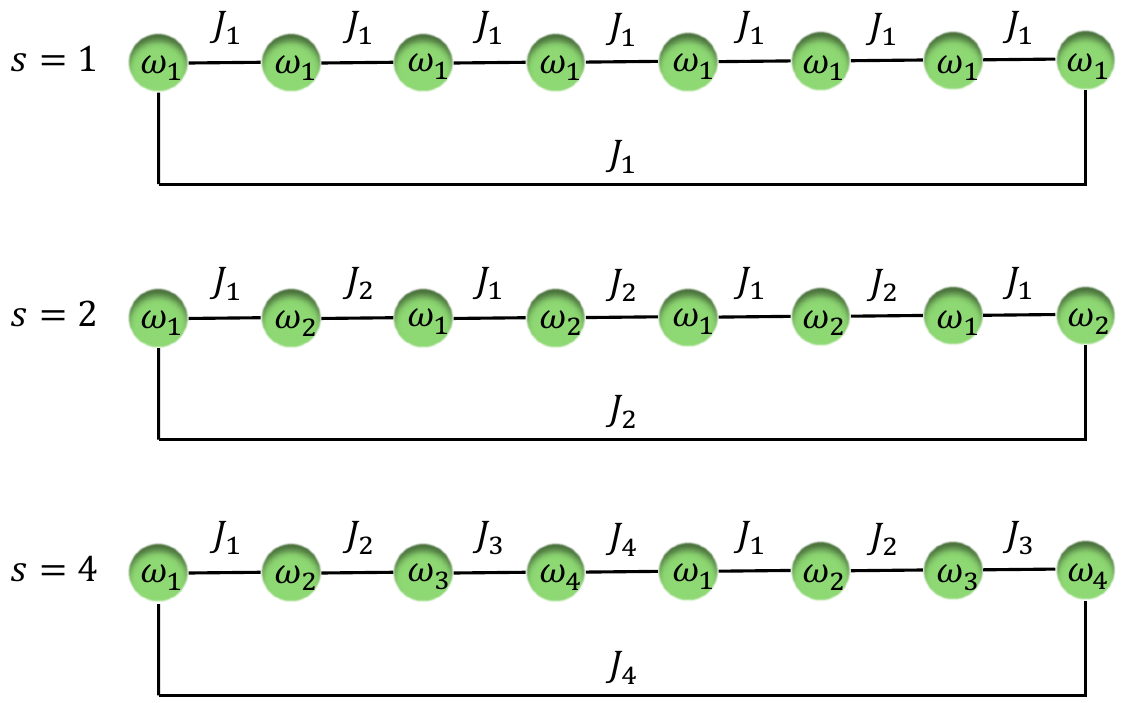}
    \caption{Schematic representation of a one-dimensional spin chain system with translational symmetry parameters $s\in\{1, 2, 4\}$, showing the corresponding patterns of interaction strengths and local external field distributions.
    }\label{fig:spinchain}
\end{figure}

We conduct two numerical experiments to benchmark iPINN-HL against the DNN-HL method. 
In the first experiment, we examine the efficiency of iPINN-HL as the system size scales and as the number of parameters to be estimated increases. First, we hold the interaction strength and external field uniform across the system (i.e. $J_{ij}=J$ and $\omega_i=\omega$ for all $i$ and nearest neighbor spins considered) and tune the number of spins $N$ in the system to compare the two algorithms. The result are shown in Fig.~\ref{fig:scaling_N}. In Fig.~\ref{fig:scaling_N}, MSE, the key indicator of the accuracy of estimation, scales polynomially with the number of input queries ${\cal N}(\{x\})$ as $\mathcal{N}^{-\ell}(\{x\})$ for both iPINN-HL and DNN-HL and the scaling coefficients $\ell$ are shown in the legend. It is clearly seen that $\ell$ for iPINN-HL is very close to the Heisenberg limit ($\ell=2$) when $N=4,7$ with only a slight decrease for $N=10$. On the other hand, $\ell$ for DNN-HL are lower than the previous case by $\sim10-15\%$. This is a clear sign showing the advantage of iPINN-HL over DNN-HL.


Next, we allow $J_{ij}$ and $\omega_i$ to take distinct values while respecting the periodic boundary condition. Namely,
\begin{equation}
	(\hat{T}^s)^{-1}H\hat{T}^s=H,
\end{equation}
where $T$ is the translation operator mapping $i$ to $i+1$ or $N$ to 1 while $s$ is a parameter indicating the extent to which
the translational symmetry is implemented. In this section, we fix $N=8$ and the cases with $s=1,2,4$ are exemplified in Fig.~\ref{fig:spinchain}. One notes that different $J_{ij}$ values as indicated in Fig.~\ref{fig:spinchain} implies the complexity of HL, with the case of $s=4$ having the maximal complexity with 8 distinct parameters ($J_{1}$ through $J_{4}$ and $\omega_{1}$ through $\omega_{4}$), the case of $s=1$ the minimal complexity with 2 distinct parameters ($J_{1}$ and $\omega_{1}$), and the case of $s=2$ in between with 4 distinct parameters  ($J_{1,2}$ and $\omega_{1,2}$).

\begin{figure*}[t]
    \centering
    \includegraphics[width=\textwidth]{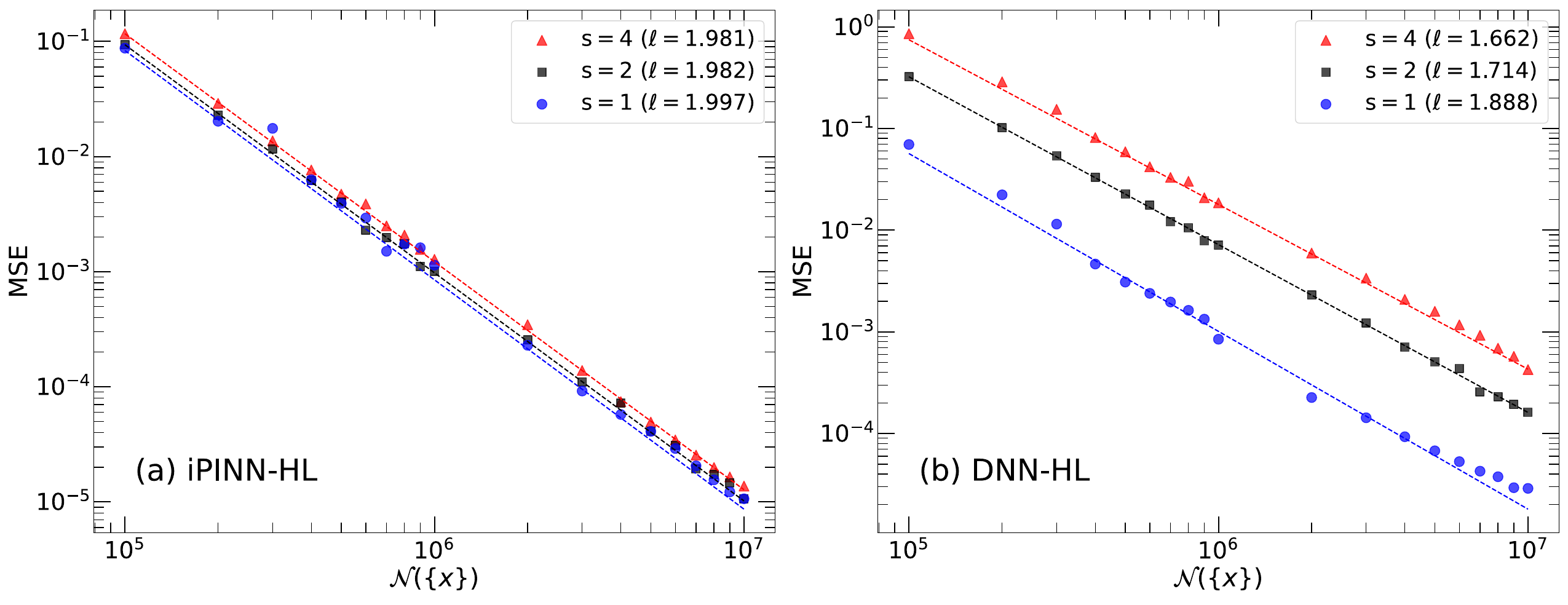}
    \caption{
    MSE scaling with the number of queries $\mathcal{N}(\{x\})$ for translation symmetry parameters $s = {1, 2, 4}$ in a one-dimensional spin chain, calculated from (a) iPINN-HL and (b) DNN-HL methods. Here the number of \blue{collocation points} $P$ is set as $50$. For $s=1$, we set $J=1$ and $\omega=0.5$. For $s=2$, $J_1=1, J_2=0.5$ and $\omega_1=1, \omega_2=0.5$. For $s=4$, $J_1=1, J_2=0.5, J_3=2, J_4=1.5$ and $\omega_1=1, \omega_2=0.5, \omega_3=1.5, \omega_4=2$.}\label{fig:scaling_s}
\end{figure*}

Again, we calculate MSE as functions of number of queries $x=(U,t,M)$, and the results
are shown in  Fig.~\ref{fig:scaling_s} with the fitted scaling coefficient $\ell$ indicated in the legend.
We see that $\ell$ has values between 1.981 and 1.997 for iPINN-HL, which is higher than the results for DNN-HL (which value between 1.662 and 1.888). We further note that for $s=1$, $\ell$ for iPINN-HL is about 5\% higher than its counterpart DNN-HL, while this percentage increases to 19\% for $s=4$. This result suggests that iPINN-HL is more efficient than DNN-HL in the cases where the Hamiltonian has higher complexity.



		\begin{table}[tb]
			\centering
			\begin{tabular}{|c|c|c|c|}
			\hline
			\#$(U,M)$ & $\Delta t=2\times 10^{-1}$ & $\Delta t=2\times 10^{-2}$ & $\Delta t=2\times 10^{-3}$ \\ \hline
			$1\times 10^4$ & $7.6636\times 10^{-2}$ & $1.4119\times 10^{-2}$ & $1.2756\times 10^{-2}$\\ \hline
			$1\times 10^5$ & $9.0494\times 10^{-4}$ & $2.9375\times 10^{-4}$ & $2.0281\times 10^{-4}$\\ \hline
			$1\times 10^6$ & $9.7906\times 10^{-6}$ & $3.4510\times 10^{-6}$ & $2.9611\times 10^{-6}$\\ \hline
			\end{tabular}
		\caption{The accuracy of estimation MSE of iPINN-HL in relation to differences $\Delta t$ of controlling parameter $t$ in data collection under fixed number of $\#(U,M)$. Here we set $P=50$ and $N=8, s=4$ with $J_1=1, J_2=0.5, J_3=2, J_4=1.5$ and $\omega_1=1, \omega_2=0.5, \omega_3=1.5, \omega_4=2$.}
		\label{tab:tab_samp_PINN}
		\end{table}
		
		\begin{table}[tb]
			\centering
			\begin{tabular}{|c|c|c|c|}
			\hline
			\#$(U,M)$ & $\Delta t=2\times 10^{-1}$ & $\Delta t=2\times 10^{-2}$ & $\Delta t=2\times 10^{-3}$ \\ \hline
			$1\times 10^4$ & $3.1211\times 10^{-1}$ & $3.0561\times 10^{-1}$ & $2.7148\times 10^{-1}$\\ \hline
			$1\times 10^5$ & $6.8932\times 10^{-3}$ & $6.3691\times 10^{-3}$ & $5.4221\times 10^{-3}$\\ \hline
			$1\times 10^6$ & $1.4892\times 10^{-4}$ & $1.0104\times 10^{-4}$ & $9.0231 \times 10^{-5}$\\ \hline
			\end{tabular}
		\caption{The accuracy of estimation MSE of DNN-HL in relation to differences $\Delta t$ of controlling parameter $t$ in data collection under fixed number of $\#(U,M)$. Here we set $N=8, s=4$ with $J_1=1, J_2=0.5, J_3=2, J_4=1.5$ and $\omega_1=1, \omega_2=0.5, \omega_3=1.5, \omega_4=2$.}
		\label{tab:tab_samp_DNN-HL}
		\end{table}

In the above study, we have shown that by incorporating physics loss, the burden of massive data collection task can be alleviated because \textit{priori} physics knowledge is embedded in the learning process. The \emph{physics loss} $\mathcal{L}_{\mathrm{physics}}$ enforces that NNQS must satisfy the Schr\"{o}dinger equation at $P$ points as shown in Eq.~\eqref{eq:PINN_phy}. In the second numerical experiment, we evaluate how $\mathcal{L}_{\mathrm{physics}}$ enhances algorithmic efficiency by analyzing the relationship between MSE and the number of queries $\mathcal{N}(\{x\})$. In the remainder of this subsection, we take $N=8$, $s=4$ with $J_1=1, J_2=0.5, J_3=2, J_4=1.5$ and $\omega_1=1, \omega_2=0.5, \omega_3=1.5, \omega_4=2$.
We fix the number of $(U,M)$ and vary the number of sampled points along the evolution trajectory of quantum state by changing the difference $\Delta t$ in the controlling parameter $t$. A smaller $\Delta t$ implies more data points in $t$, while a larger $\Delta t$ suggests otherwise.
The result can be found in  Table~\ref{tab:tab_samp_PINN} and  Table~\ref{tab:tab_samp_DNN-HL} respectively for iPINN-HL and DNN-HL. Overall the MSE values in  Table~\ref{tab:tab_samp_PINN} are at least one order of magnitude smaller than those in Table~\ref{tab:tab_samp_DNN-HL}, confirming the superiority of iPINN-HL method. For iPINN-HL, reducing $\Delta t$ from $2 \times 10^{-1}$ to $2 \times 10^{-2}$ decreases the MSE by a factor of 3-4. However, further reducing $\Delta t$ to $2 \times 10^{-3}$ yields only a marginal improvement (less than a factor of 1.5). This suggests the existence of an optimal sampling density--beyond which additional queries provide diminishing returns in reconstruction accuracy for the HL problem. On the other hand, this phenomenon is not observed in the DNN-HL results (Table~\ref{tab:tab_samp_DNN-HL}), where the MSE shows no consistent improvement with decreasing $\Delta t$.
This insensitivity to temporal resolution suggests that DNN-HL lacks the physics-informed regularization to achieve convergence with optimal sampling efficiency. The ability of iPINN-HL to extract meaningful physical constraints from sparser data compared to conventional deep learning approaches is therefore a key advantage.

\begin{figure}
	\centering
    \includegraphics[width=\linewidth]{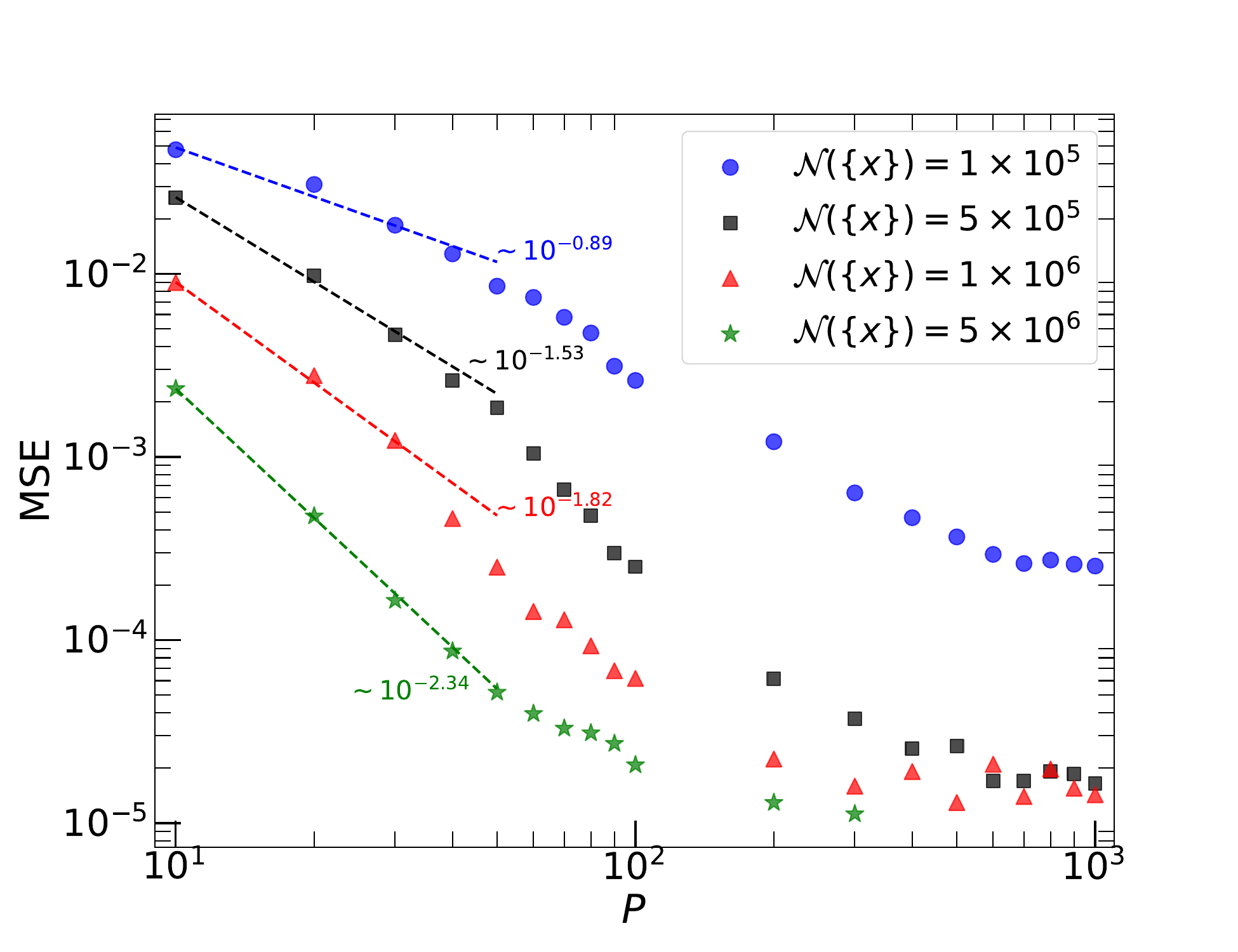}
    \caption{
    Scaling of the MSE with the number of \blue{collocation points} $P$ for different number of queries $\mathcal{N}(\{x\})$. Dashed lines show fitted slopes of the initial polynomial decay regime with exponents indicated.  Here we set $P=50$ and take $N=8, s=4$ with $J_1=1, J_2=0.5, J_3=2, J_4=1.5$ and $\omega_1=1, \omega_2=0.5, \omega_3=1.5, \omega_4=2$.}\label{fig:scaling_P}
\end{figure}

Furthermore, we examine the performance of iPINN-HL when the number of \blue{collocation points} $P$ is varied. With larger $P$, the NNQS must satisfy the Schr\"{o}dinger equation at more points and receives greater penalty from $\mathcal{L}_{\mathrm{physics}}$ if it deviates from the prediction from the Schr\"{o}dinger equation. The result can be found in Fig.~\ref{fig:scaling_P}. 
\blue{Fig.}~\ref{fig:scaling_P} reveals that the MSE initially follows a polynomial decay with increasing $P$, where the scaling exponent grows with the number of queries $\mathcal{N}(\{x\})$. This demonstrates a synergistic relationship between data availability and the efficacy of the physics-informed loss $\mathcal{L}_{\mathrm{physics}}$. Notably, when $P$ exceeds a critical threshold ($P \approx 100$), the MSE saturates--particularly for small $\mathcal{N}(\{x\})$--indicating a fundamental limit to physics-based enhancement without sufficient supporting data. These results highlight the complementary roles of data-driven and physics-driven learning: while $\mathcal{L}_{\mathrm{physics}}$ significantly reduces the demand for extensive training data (compared to purely data-driven methods like DNN-HL), optimal performance requires a balanced integration of both $\mathcal{L}_{\mathrm{data}}$ and $\mathcal{L}_{\mathrm{physics}}$ components. This hybrid approach enables high-accuracy solutions with substantially fewer data points than conventional neural networks. \blue{To further elucidate the MSE saturation observed in Fig.~\ref{fig:scaling_P}, we consider an extremal case of no observational data ($\mathcal{N}(\{x\}) = 0$), minimizing $\mathcal{L}_{\mathrm{physics}}$ alone permits convergence to any possible NNQS that satisfies the Schr\"{o}dinger equation under some Hamiltonian, yielding a zero physics loss while the estimate $\bm{\hat{\theta}}$ can take any value, hence physics loss alone provides no contribution without data loss. Increasing $P$ beyond a threshold provides no further constraint without data, leading to such saturation. Similar phenomenon is also observed when solving the motion of a gravity pendulum using framework of PINN \cite{lu2021learning}.}


\subsection{Cross-Resonance Gates}\label{sec:CR_gate}


We employ  iPINN-HL  to reconstruct and calibrate the Hamiltonian of a quantum system implementing a cross-resonance (CR) gate. This challenging learning task is complicated by the inherent noise characteristics of multi-qubit gates. 

The CR gate  serves as a fundamental two-qubit entangling gate in fixed-frequency superconducting transmon architectures, requiring only microwave control. Through appropriate pulse sequences, the CR gate can be converted into a locally equivalent CNOT gate, making it particularly valuable for quantum circuit compilation. The system Hamiltonian governing the CR gate dynamics is given by:
\begin{equation}
	H_\mathrm{CR}=\frac{\sigma_z\otimes A}{2}+\frac{I\otimes B}{2},
\end{equation} 
with $A = c_{zi}I+c_{zx}\sigma_x+c_{zy}\sigma_y+c_{zz}\sigma_z$ and $B = c_{ix}\sigma_x+c_{iy}\sigma_y+c_{iz}\sigma_z$, where $\sigma_x$, $\sigma_y$ and $\sigma_z$ are the Pauli matrices and $I$ is a $2\times2$ identity matrix. In this sense, HL for the CR gate reduces to estimating the parameter vector $\bm{c}=\{c_{zi},c_{zx},c_{zy},c_{zz},c_{ix},c_{iy},c_{iz}\}$. Importantly, we emphasize that throughout this section and Section~\ref{sec:Crosstalk}, our protocol operates under two key constraints: (1) no preparation of entangled states is permitted, and (2) measurements are restricted to product-state projections only. These operational restrictions reflect realistic limitations in many superconducting quantum processors while still enabling accurate Hamiltonian characterization. 




The calibration of CR gates is particularly susceptible to noise in realistic quantum systems. In this work, we focus on two dominant noise sources: readout noise and decoherence, which represent the most significant limitations for Hamiltonian parameter estimation.

\begin{itemize}
	\item \textit{Readout noise.} 
	Readout noise introduces errors during quantum state measurement, corrupting experimental data and biasing the inferred Hamiltonian parameters. This directly affects the reliability of gate calibration.

The readout line of a qubit measurement result operates as a classical channel, where bit-flip errors can occur. Specifically, the true measurement outcome $y \in \{0,1\}$ may be flipped. We denote the noisy measurement outcome observed in the experiment as $\tilde{y}$. The readout noise can be modeled by the conditional probability $p_{\tilde{y}|y}(\tilde{y}|y)$. In the absence of readout noise, we always have $p_{\tilde{y}|y}(\tilde{y}=y|y)=1$. Therefore, the probability of obtaining the noisy outcome $\tilde{y}_k$ in the $k$-th entry of the dataset $D$ is given by:
	\begin{equation}
	\begin{aligned} \label{eq:biflip}
		p(\tilde{y}_{ki}|x_k,\bm{\theta})&=p_{\tilde{y}_{ki}|y_{ki}}(\tilde{y}_{ki}|\tilde{y}_{ki})p(y_{ki}|x_k,\bm{\theta})\\
		&+p_{\tilde{y}_{ki}|y_{ki}}(\tilde{y}_{ki}|1-\tilde{y}_{ki})p(y_{ki}|x_k,\bm{\theta}),
	\end{aligned}
	\end{equation}
where $y_{ki}$ and $\tilde{y}{ki}$ denote the true and noisy measurement outcomes, respectively, of the $i$-th qubit in the $k$-th entry of the dataset $D$, and $p(y|x,\bm{\theta})$ is defined in Eq.~\eqref{eq:dataloss_DNN-HL}. In this study, we assume that the bit-flip channel is symmetric, i.e., $p_{\tilde{y}|y}(1|0) = p_{\tilde{y}|y}(0|1)$, and that the bit-flip channel is identical for all qubits in the system for simplicity.
	\item \textit{Decoherence.} 
	Decoherence degrades quantum coherence, leading to deviations from ideal unitary dynamics $U(t;\theta) = e^{-iH(\theta)t}$ and a reduction in gate fidelity. Here, we modeled it as a completely depolarized channel:
		\begin{equation}\label{eq:decoherence}
			\mathcal{E}(\rho(t))=(1-p_d(t))\rho(t)+p_d(t)\frac{\mathcal{I}}{2^n},
		\end{equation}
		where $\rho(t)=U(t;\theta)\rho(0)U^\dag(t;\theta)$ and $\mathcal{I}$ is a $4\times4$ matrix. To first order, the occurrence of depolarization events can be modeled as a Poisson process with rate  $\mu$:
		\begin{equation}
			p_d(t)=\mathrm{exp}\left(-\frac{t-t_0}{\mu}\right),
		\end{equation}
		where $t_0$ is the starting time of the experiment. Therefore, identifying and quantifying decoherence is equivalent to estimating the Poisson rate $\mu$. \blue{Same model is used in Ref.~\cite{dutt2023active}. This approach approximates a Lindblad master equation with strong dephasing, mimicking rapid decoherence toward a maximally mixed state, and we include a full derivation in Appendix~\ref{appendix.b} to justify this approximation. Using this model for decoherence, we avoid making modification to the neural network to represent density matrix, which is harder to optimize when system scales up.}
\end{itemize}

To summarize, the calibration of CR gate in the presence of noise is reduced to estimate the vector $\bm{c}$, bit-flip rate $p_{\tilde{y}|y}(\tilde{y}|y)$ and Poisson rate $\mu$ from the experiment data. 


The relationship between estimation accuracy, measured by MSE, and the number of queries is illustrated in Fig.~\ref{fig:scaling_noise}. It is evident from the figure that iPINN-HL consistently outperforms DNN-HL across both noiseless and noisy conditions. Notably, the MSE for iPINN-HL declines more rapidly with increasing query numbers, demonstrating its superior scalability and robustness. Noise sources, particularly readout noise, have a pronounced impact on the calibration process, as seen by the more gradual decline and greater spread in the DNN-HL curves compared to the iPINN-HL curves. Moreover, the presence of noise also affects training stability, as reflected in the less smooth scaling behavior of MSE for DNN-HL. These results confirm that iPINN-HL offers greater resilience against noise and is more efficient in terms of query usage compared to the traditional DNN-HL approach.
\blue{The final Relative MSE for the two algorithms under noisy and noiseless setting is shown in TABLE~\ref{tab:RMSE}.
\begin{table}[!htb]
\centering
\caption{Relative MSE for iPINN-HL and DNN-HL under noiseless noisy settings, $\mathcal{N}(\{x\})=10^5$.}
\label{tab:RMSE}
\begin{tabular}{|c|c|c|}
\hline
Method & iPINN-HL & DNN-HL \\
\hline
noiseless & $1.1 \times 10^{-2}$ & $1.9\times 10^{-2}$\\
noisy & $1.8\times 10^{-2}$ & $2.8\times 10^{-2}$ \\
\hline
\end{tabular}
\end{table}

\begin{figure}
	\centering
    \includegraphics[width=\linewidth]{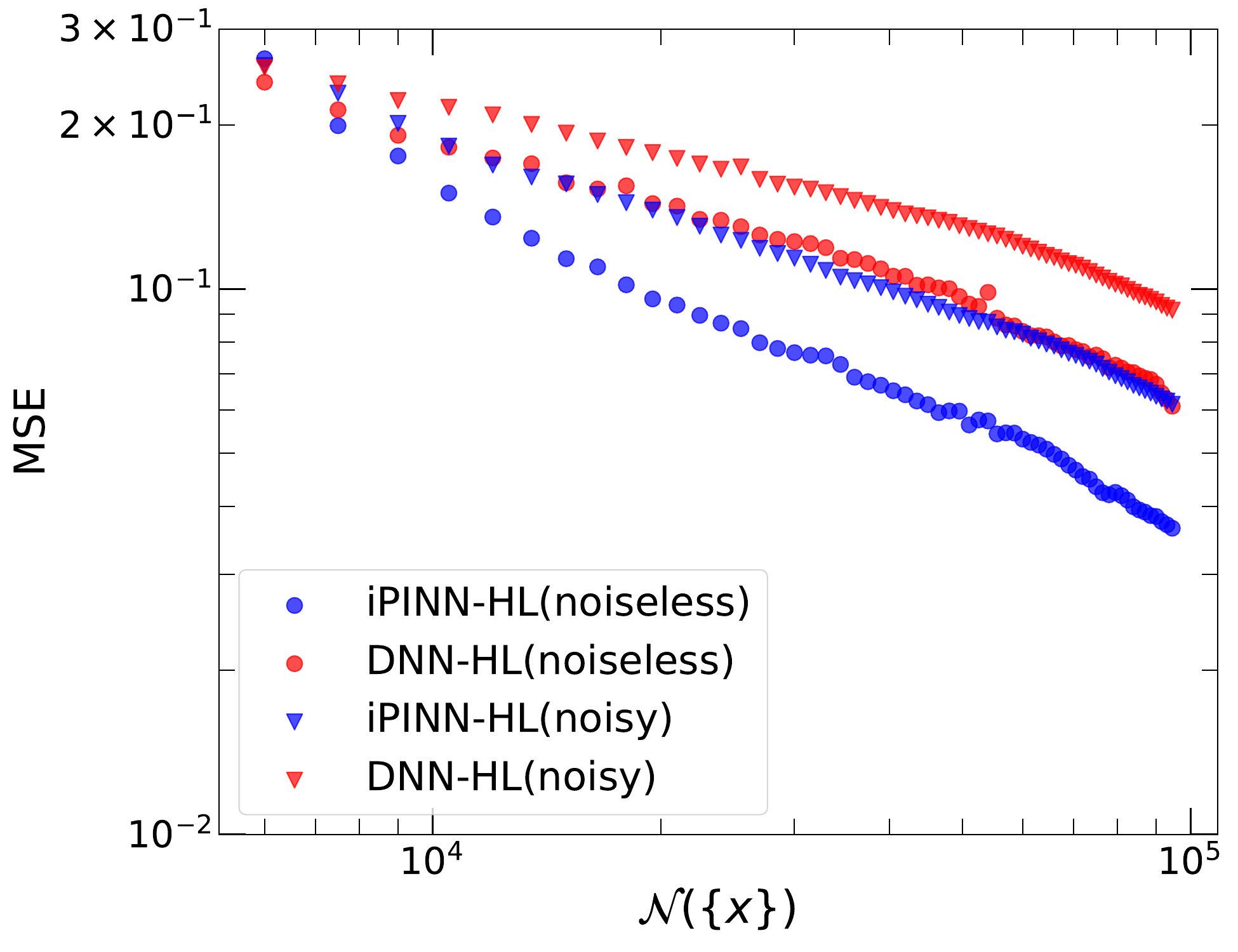}
    \caption{Calibration accuracy of CR gate MSE as a function of the number of queries $x$ in the dataset $D$, comparing iPINN-HL and DNN-HL methods under both noiseless and noisy conditions. Here we set the parameter vector $\bm{c}=\{c_{zi},c_{zx},c_{zy},c_{zz},c_{ix},c_{iy},c_{iz}\}=\{0.5,0.6,0.7,0.8,0.9,1,1.1\}$, Poisson rate $\mu=5$ and the bit-flip error rate as $0.995$ in the readout noise. 
    }\label{fig:scaling_noise}
\end{figure}
}

\begin{table*}[!htb]
    
    \begin{minipage}{0.5\textwidth}
      \centering
        \begin{tabular}{|c|c|c|c|c|c|}
        \hline
           & Qubit 1 & Qubit 2 & Qubit 3 & Qubit 4 \\ \hline
           Qubit 1 & 0.00 & 0.59 & 0.46 & 0.40 	\\ \hline
           Qubit 2 & 0.59 & 0.00 & 0.35 & 0.48 \\ \hline
           Qubit 3 & 0.46 & 0.35 & 0.00 & 0.54 \\ \hline
           Qubit 4 & 0.40 & 0.48 & 0.54 & 0.00 \\ \hline
        \end{tabular}
    \end{minipage}%
    \begin{minipage}{0.5\textwidth}
      \centering
        \begin{tabular}{|c|c|c|c|c|c|}
        \hline
           & Qubit 1 & Qubit 2 & Qubit 3 & Qubit 4 \\ \hline
           Qubit 1 & 0.00 & 0.25 & 0.37 & 0.23 	\\ \hline
           Qubit 2 & 0.25 & 0.00 & 0.56 & 0.67 \\ \hline
           Qubit 3 & 0.37 & 0.56 & 0.00 & 0.27 \\ \hline
           Qubit 4 & 0.23 & 0.67 & 0.27 & 0.00 \\ \hline
        \end{tabular}

    \end{minipage} 
    \caption{The table of crosstalk coefficients in the system with $XY$-type crosstalk $(\eta_{ij})$  shown in the left panel and $ZZ$-type  $(\epsilon_{ij})$ crosstalk in the right panel.}
    \label{tab:crosstalk}
\end{table*}

\subsection{Crosstalks}\label{sec:Crosstalk}
In this subsection, we focus on the HL problem for crosstalk effects in multi-qubit systems. Crosstalk refers to unintended interactions between qubits in a quantum system that occur due to imperfections in hardware or control signals. For a system with $N$ qubits, the general crosstalk effect can be modeled as:


\begin{equation}
	\begin{aligned}
		H_\mathrm{crosstalk}&=\sum_{i<j}\eta_{ij}(\sigma_x^{(i)}\sigma_x^{(j)}+\sigma_y^{(i)}\sigma_y^{(j)})+\sum_{i<j}\epsilon_{ij}\sigma_z^{(i)}\sigma_z^{(j)}\\
		&+\sum_i \delta_i \sigma_z^{(i)}.
	\end{aligned}
\end{equation}
The first term is caused by unintended $XY$-type interactions, where qubits influence each other through off-diagonal terms in the Hamiltonian. The second term is $ZZ$-type interactions, which can arise from dephasing-induced crosstalk when two qubits are affected by shared environmental noise sources or from residual coupling between two qubits. The last term is sometimes called ``classical crosstalk" \cite{wang2022control} since it is merely a shift in the qubit frequency and not qubit-qubit interaction. This can occur when qubits are not sufficiently detuned from one another, causing their frequencies to overlap and leading to interference \cite{hertzberg2021laser}. This can also occur when there is drive-pulse crosstalk \cite{yang2024mitigation}.

Since qubits can be more appropriately detuned from each other with advancing quantum technology and the drive-pulse crosstalk can be readily characterized by measuring Rabi frequency, we focus on quantum crosstalk in this study:
\begin{equation}
	H_\mathrm{crosstalk}=\sum_{i<j}\eta_{ij}(\sigma_x^{(i)}\sigma_x^{(j)}+\sigma_y^{(i)}\sigma_y^{(j)})+\sum_{i<j}\epsilon_{ij}\sigma_z^{(i)}\sigma_z^{(j)}
\end{equation}
Furthermore, in our query $x=(U,t,M)$ to the system, we assume that the initial preparation unitary $U$ is perfect,  and projective measurement onto product state is also perfect because single qubit gate can be fine-calibrated using dynamical decoupling to cancel out the interaction in the process. Readout error and decoherence is not included in the model for simplicity but can also be estimated in principle.

\begin{figure*}[t]
    \centering
    \includegraphics[width=\textwidth]{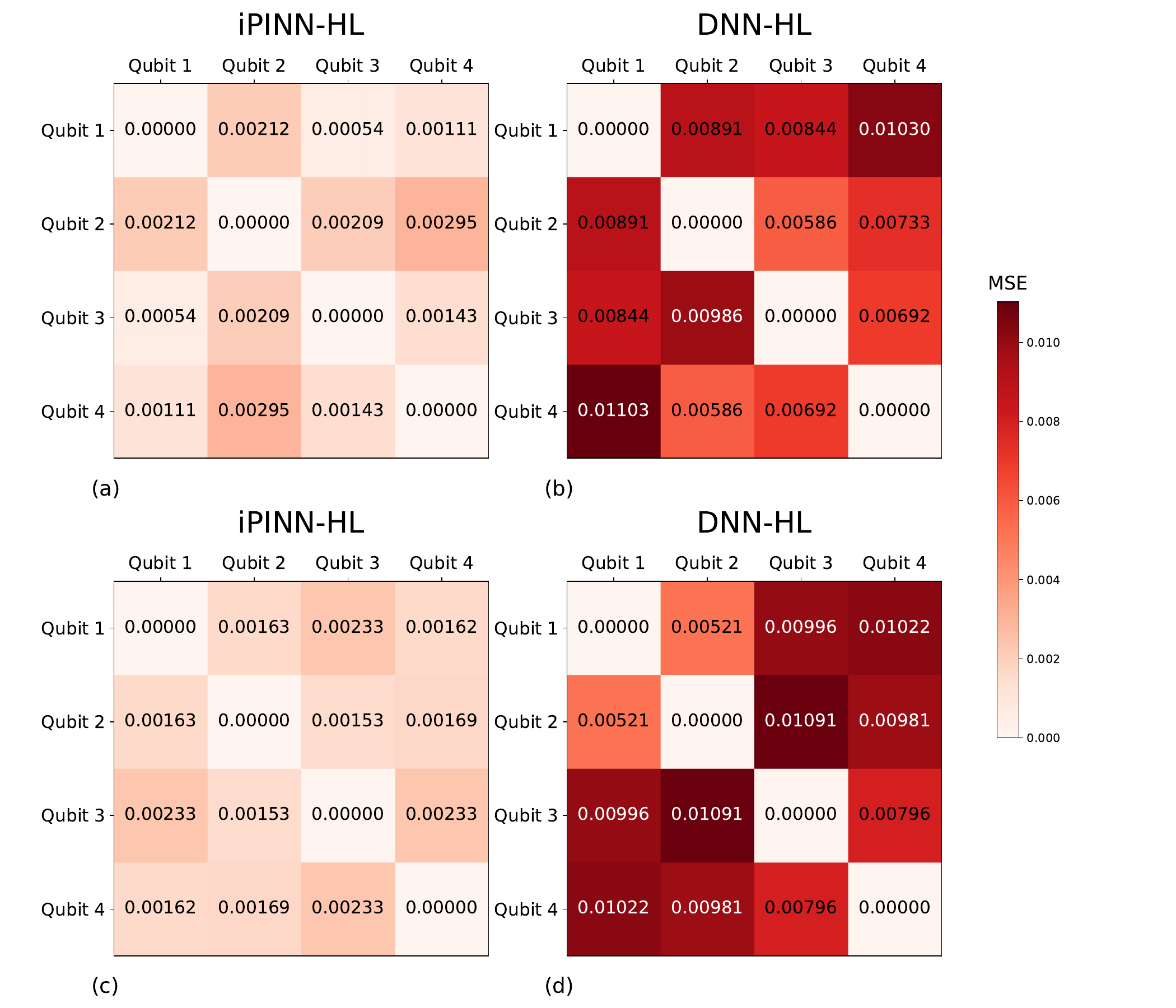}
    \caption{The accuracy of estimation MSE (shown as pseudo-color scale) of coefficient for $XY$- and $ZZ$-type crosstalk using iPINN-HL in (a) and (c). And same for (b) and (d) using DNN-HL. The MSE is averaged based on multiple different initialization of algorithms. The crosstalk coefficients are set according to Table~\ref{tab:crosstalk}.
    }\label{fig:crosstalk_MSE}
\end{figure*}

Here, we conduct a numerical study to estimate the crosstalk of a four-qubit array. We set the parameters $\eta_{ij}$ of $XY$-type crosstalk and $\epsilon_{ij}$ of $ZZ$-crosstalk according to Table~\ref{tab:crosstalk}.  
  The accuracy of the parameter estimation, measured by MSE, is presented in Fig.~\ref{fig:crosstalk_MSE}. The figure clearly illustrates that iPINN-HL consistently achieves a significantly lower MSE across all crosstalk parameters compared to DNN-HL.
Specifically, for both $XY$- and $ZZ$-type crosstalk, the iPINN-HL method exhibits MSE values that are approximately $5$ to $10$ times smaller than those of DNN-HL. This superior performance holds consistently across different parameter initializations and system configurations, as seen from the stable, low values  of the results from iPINN-HL in the left column. Additionally, the MSE of DNN-HL displays greater fluctuations, indicating its sensitivity to initialization and its lower robustness when faced with complex crosstalk effects.These results highlight iPINN-HL's substantial advantage in both accuracy and stability when characterizing multi-qubit systems affected by crosstalk, reinforcing its potential for practical deployment in quantum device calibration and control.

\subsection{Compensation for Parameter Drifts}\label{sec:drift}
We extend  iPINN-HL  to address the problem of parameter drifting in quantum systems, where Hamiltonian parameters change over time due to environmental fluctuations or system instabilities. External factors such as temperature fluctuations, electromagnetic noise, and mechanical vibrations can subtly alter the system dynamics. Internally, qubit decoherence, residual interactions, and imperfections in hardware calibration contribute to this drift. Additionally, aging of materials or components in the device can cause long-term changes in performance. 

In this section, we employ both iPINN-HL and DNN-HL to capture the parameter shift and re-calibrate the system, thereby compensating for the effects of parameter drift. 
After training the iPINN-HL and DNN-HL algorithms to convergence, we introduce a sudden change to the system parameters. Although such abrupt changes are uncommon in real experimental settings, they serve as an effective testbed to assess the adaptability and robustness of both methods. Following this parameter shift, we collect a new batch of data $\Delta \mathcal{N}(\{x\})$ to refine the parameter estimates. \blue{$\Delta \mathcal{N}(\{x\})$ contains 300 queries as shown in Fig.~\ref{fig:query_model} and $x=(U,t,M)$ is randomly sampled.} For both algorithms, the pre-trained network parameters are retained as the initial model configuration, and the networks are further trained using the new dataset $\Delta \mathcal{N}(\{x\})$. This approach enables the models to efficiently adapt to the updated system dynamics by integrating the latest observational data.



This approach closely aligns with the online learning paradigm in ML \cite{marschall2020unified}, where models incrementally adapt as new data becomes available, rather than being retrained from scratch. This way, iPINN-HL benefits from its pre-trained structure, enabling it to rapidly converge to the new parameter regime after a sudden change. In contrast, DNN-HL requires more iterations and data to stabilize its estimates, as it lacks the inherent incorporation of governing physical principles that facilitate efficient adaptation.

\begin{figure}
	\centering
    \includegraphics[width=\linewidth]{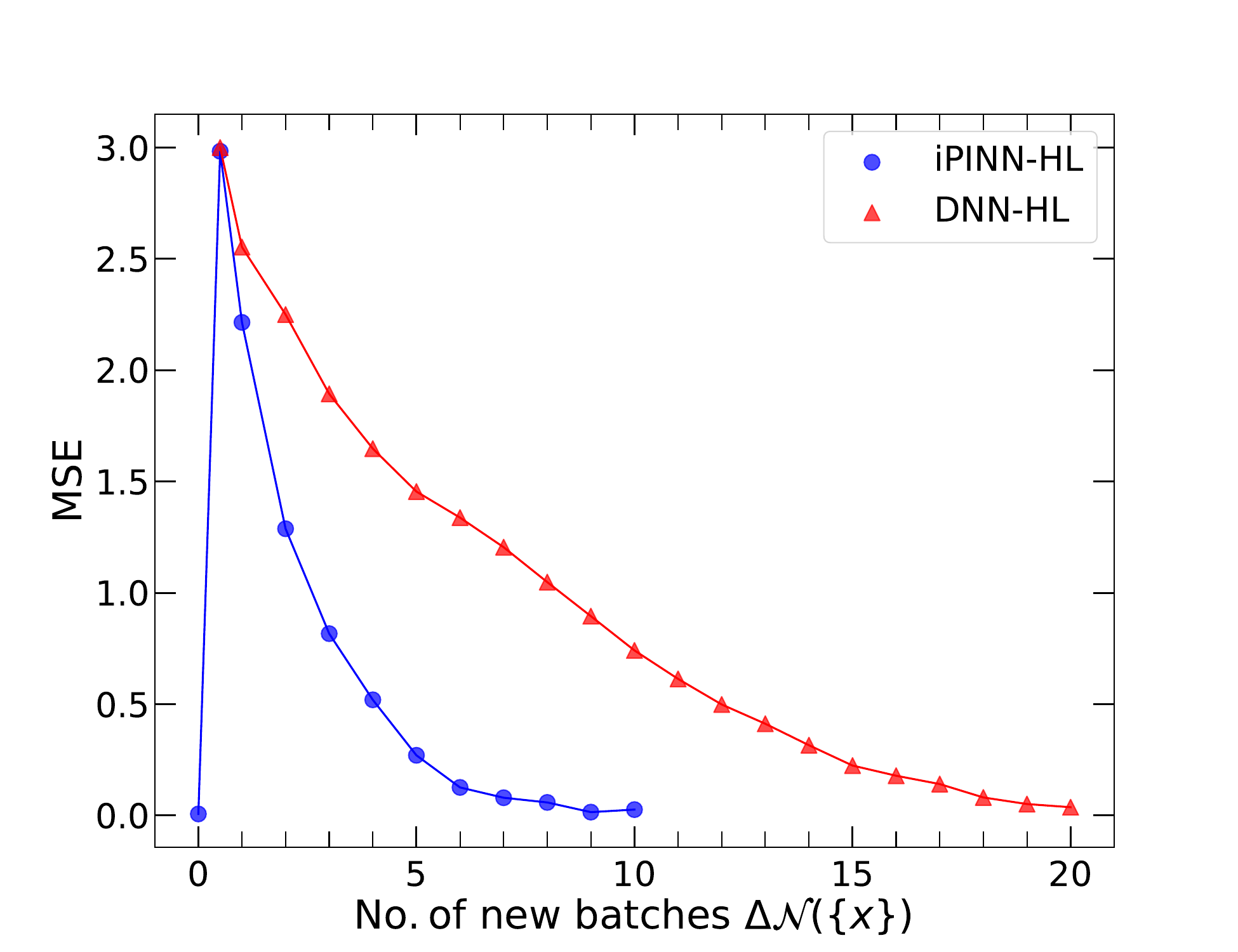}
    \caption{The accuracy of estimation MSE of iPINN-HL and DNN-HL after a sudden change in the parameters of a two-qubit Hamiltonian defined in \blue{Eq.~(\ref{eq:drift})}. \blue{Every new batch $\Delta \mathcal{N}(\{x\})$ contains 300 queries depicted in Fig.~\ref{fig:query_model}.}
    }\label{fig:online}
\end{figure}

For this study, we use the following two-qubit Hamiltonian:
\begin{equation}
	H=\omega_1\sigma_z^{(1)}+\omega_2\sigma_z^{(2)}+\varepsilon\sigma_z^{(1)}\sigma_z^{(2)}, \label{eq:drift}
\end{equation}
where initially, $\omega_1=\omega_2=0.5$ and $\varepsilon=1$. After the sudden change, the parameters shift to $\omega_1=\omega_2=1.5$ and $\varepsilon=2$. In the numerical experiment, the size of the newly collected queries $\Delta \mathcal{N}(\{x\})$ is $300$. The results, shown in Fig.~\ref{fig:online}, indicate that MSE for iPINN-HL quickly converges to zero after just $8$ batches of $\Delta \mathcal{N}(\{x\})$, while the MSE for DNN-HL decreases much more slowly and remains above $1$ even after $10$ batches. \blue{DNN-HL requires twice the amount of data to reach same level of accuracy again with iPINN-HL after parameter drift.} This highlights the superior adaptability and convergence efficiency of iPINN-HL in response to abrupt system parameter changes.

\section{Conclusion and Outlook} \label{sec:conclusion}

In this study, we presented iPINN-HL, a physics-informed neural network framework for HL. Through comprehensive numerical experiments across multiple scenarios---including one-dimensional spin chain systems, cross-resonance gate calibration, crosstalk identification, and parameter drift compensation---we demonstrated its superior performance compared to DNN-HL. Our results, as detailed in Sec.~\ref{sec:results}, show that iPINN-HL achieves higher estimation accuracy, enhanced robustness to noise, and greater adaptability in response to system changes.

The key advantage of iPINN-HL lies in its integration of physical principles into the learning process, enabling it to approach near-Heisenberg limit performance and maintain accuracy even in noisy environments. Its ability to adapt to dynamic system parameters makes it particularly suitable for real-time quantum system characterization and calibration.

Moreover, iPINN-HL is highly compatible with advanced ML techniques. For instance, it can be integrated as a subroutine within other ML-based HL protocols, as demonstrated in prior works \cite{flurin2020using, tucker2024hamiltonian}. Additionally, iPINN-HL has the potential to be enhanced through active learning strategies \cite{dutt2023active}, allowing for more efficient resource utilization while maintaining high accuracy. These features position iPINN-HL as a powerful and flexible tool for advancing HL in both research and practical quantum technology applications.

Future developments involving the integration of iPINN-HL with experimental quantum platforms hold considerable promise for enhancing the precision and reliability of quantum device calibration and control. Research directions include scaling iPINN-HL to larger and more complex quantum systems and devising hybrid strategies that combine classical optimization with quantum-assisted Hamiltonian learning. Extending iPINN-HL to open quantum systems with non-unitary dynamics also presents a compelling challenge, with the potential to bridge theoretical advances in Hamiltonian learning and practical applications in quantum technologies.

\section*{ACKNOWLEDGEMENT}

This work is supported by the National Natural Science Foundation of China (Grant Nos. 11874312 and 12474489), the Research Grants Council of Hong Kong (CityU 11304920), Shenzhen Fundamental Research Program (Grant No. JCYJ20240813153139050), the Guangdong Provincial Quantum Science Strategic Initiative (Grant Nos. GDZX2203001, GDZX2403001), and the Innovation Program for Quantum Science and Technology (Grant No. 2021ZD0302300). The calculations involved in this work were partially performed on the Tianhe Xingyi supercomputer at the National Supercomputer Center in Guangzhou, China. Open Access made possible with partial support from the Open Access Publishing Fund of the City University of Hong Kong.

\section*{DATA AVAILABILITY}
The data that support the findings of this article, \blue{along with the source codes}, are openly available \cite{data}. \blue{Some training details of iPINN-HL and DNN-HL are also provided in Appendix~\ref{sec:training_details}.}

\appendix
\blue{
\section{Baseline Comparison}

In this appendix, we provide a comparison of several baseline methods for quantum state tomography, focusing on their scalability, computational cost, and accuracy. The methods considered are:

\begin{itemize}
    \item \textbf{iPINN-HL} (Physics-Informed Neural Networks for Hamiltonian Learning): The model proposed in this paper, where physics is incorporated into the learning process through a physics-informed loss term.
    \item \textbf{DNN-HL} (Deep Neural Network for Hamiltonian Learning): A traditional deep neural network approach to learning Hamiltonian parameters, without the inclusion of physics-informed constraints.
   \item \textbf{RBM-HL} (Restricted Boltzmann Machine for Hamiltonian Learning):
In the RBM-HL method, we replace the neural network architecture in DNN-HL with an RBM. 
	\item \textbf{MLE-HL}(Maximum Likelihood Estimation): A traditional, non-ML approach used for quantum tomography and HL.
\end{itemize}

We confine this comparison to the one-dimensional spin chain model described in Sec.~\ref{sec:spinchain}, as it enables a controlled evaluation of the scaling behavior, computational cost, and accuracy of each method under varying system sizes and numbers of parameters. This well-understood and computationally feasible model provides an ideal framework for assessing the core performance characteristics of the methods. 

\subsection{RBM-HL}
\paragraph{RBM representation of quantum state.} 
The RBM consists of a visible layer $ \mathbf{v} $ and a hidden layer $ \mathbf{h} $. 
We denote the visible layer input as $ \mathbf{v}(t) = (m, t) $, where $ m $ is the bit-string representation of the computational basis state $ \ket{m} $, and $ t $ is the time parameter. 
The hidden layer encodes latent variables that capture correlations in the quantum system.

The energy function of the RBM is defined as
\begin{align}
    E(\mathbf{v}, \mathbf{h}; \mathbf{W}, \mathbf{b}, \mathbf{c}) 
    &= -\sum_{i} v_i c_i - \sum_{j} h_j b_j - \sum_{i,j} v_i w_{ij} h_j ,
\end{align}
where $ \mathbf{v} = (m_1, m_2, \dots, m_n, t) $ represents the visible units, 
$ \mathbf{h} = (h_1, h_2, \dots, h_m) $ are the hidden units, 
$ w_{ij} $ is the weight connecting the $ i $-th visible unit and the $ j $-th hidden unit, 
and $ c_i $ and $ b_j $ are the biases for visible and hidden layers, respectively.

To model a quantum state, the RBM defines a variational wavefunction ansatz. 
The amplitude associated with a configuration $\mathbf{v}$ is
\begin{equation}
    \Psi_{\boldsymbol{\omega}}(\mathbf{v}) 
    = \sum_{\mathbf{h}} e^{-E(\mathbf{v}, \mathbf{h}; \mathbf{W}, \mathbf{b}, \mathbf{c})},
\end{equation}
where the parameters $\boldsymbol{\theta} = \{\mathbf{W}, \mathbf{b}, \mathbf{c}\}$ can be complex-valued. 

The normalized quantum state is then given by
\begin{equation}
    \ket{\Psi_{\boldsymbol{\omega}}} = 
    \frac{1}{\sqrt{\mathcal{N}}} \sum_{\mathbf{\omega}} \Psi_{\boldsymbol{\omega}}(\mathbf{v}) \ket{\mathbf{v}},
\end{equation}
with normalization constant
\begin{equation}
    \mathcal{N} = \sum_{\mathbf{v}} \big|\Psi_{\boldsymbol{\omega}}(\mathbf{v})\big|^2 .
\end{equation}

The probability of observing outcome $\mathbf{v}$ upon measurement in the computational basis follows the Born rule:
\begin{equation}
    P(\mathbf{v}) = \frac{\big|\Psi_{\boldsymbol{\omega}}(\mathbf{v})\big|^2}{\sum_{\mathbf{v'}} \big|\Psi_{\boldsymbol{\omega}}(\mathbf{v'})\big|^2}.
\end{equation}

Training the RBM consists of adjusting the weights and biases to minimize the difference between the model distribution and the measurement statistics. 
This is typically done using contrastive divergence (typically denoted as CD-$k$ \cite{Hinton2012}) in cases where the target distribution is directly accessible from experimental measurement outcomes. 

\paragraph{RBM Training from Quantum Measurements.}  
Given a set of measurement samples $\{\mathbf{v}^{(i)}\}$ drawn from the true quantum state, the RBM parameters 
$\boldsymbol{\omega} = \{\mathbf{W}, \mathbf{b}, \mathbf{c}\}$ are optimized such that the model probability 
distribution $P_{\boldsymbol{\omega}}(\mathbf{v})$ matches the empirical statistics. 
The optimization objective is the log-likelihood
\begin{equation}
    \mathcal{L}(\boldsymbol{\omega}) = \sum_{i} \log P_{\boldsymbol{\omega}}(\mathbf{v}^{(i)}).
\end{equation}
Its gradient can be written as
\begin{equation}
    \frac{\partial \mathcal{L}}{\partial \boldsymbol{\omega}}
    = \mathbb{E}_{\text{data}} \!\left[ \frac{\partial E}{\partial \boldsymbol{\omega}} \right]
    - \mathbb{E}_{\text{model}} \!\left[ \frac{\partial E}{\partial \boldsymbol{\omega}} \right],
\end{equation}
where
\begin{align}
    \mathbb{E}_{\text{data}}[f(\mathbf{v},\mathbf{h})] 
    &= \frac{1}{N} \sum_{i=1}^N \sum_{\mathbf{h}} P_{\boldsymbol{\omega}}(\mathbf{h}|\mathbf{v}^{(i)}) f(\mathbf{v}^{(i)},\mathbf{h}), \\
    \mathbb{E}_{\text{model}}[f(\mathbf{v},\mathbf{h})] 
    &= \sum_{\mathbf{v},\mathbf{h}} P_{\boldsymbol{\omega}}(\mathbf{v},\mathbf{h}) f(\mathbf{v},\mathbf{h}).
\end{align}
The first term, known as the \emph{positive phase}, is directly estimated from measurement data.  
The second term, known as the \emph{negative phase}, requires sampling from the model distribution.

The parameter updates then read
\begin{align}
    \Delta W &= \eta \big( \langle \mathbf{v}\mathbf{h}^\top \rangle_{\text{data}} - \langle \mathbf{v}\mathbf{h}^\top \rangle_{\text{model}} \big), \\
    \Delta \mathbf{c} &= \eta \big( \langle \mathbf{v} \rangle_{\text{data}} - \langle \mathbf{v} \rangle_{\text{model}} \big), \\
    \Delta \mathbf{b} &= \eta \big( \langle \mathbf{h} \rangle_{\text{data}} - \langle \mathbf{h} \rangle_{\text{model}} \big),
\end{align}
with learning rate $\eta$.

To approximate $\mathbb{E}_{\text{model}}$, one uses Gibbs sampling (i.e. CD-$k$).  
For binary RBMs, the conditional probabilities needed for Gibbs updates are
\begin{align}
    P(h_j = 1 | \mathbf{v}) &= \sigma(b_j + \mathbf{v}^\top W_{\cdot j}), \\
    P(v_i = 1 | \mathbf{h}) &= \sigma(c_i + W_{i\cdot}\mathbf{h}),
\end{align}
where $\sigma(x) = 1/(1+e^{-x})$.  
In CD-$k$, one initializes $\mathbf{v}$ with a data sample, alternates sampling $\mathbf{h}$ and $\mathbf{v}$ for $k$ steps, and uses the resulting samples to estimate the negative-phase expectations.
 
Unlike standard DNNs, which rely on direct backpropagation through deterministic forward passes, RBM training requires sampling from the model distribution to estimate the negative-phase statistics. This involves iterative Gibbs sampling or its variants, which can be computationally demanding due to the need for repeated stochastic updates and Markov chain mixing. As a result, each training step of an RBM typically incurs significantly higher computational cost compared to DNNs, especially when scaling to large systems.

\subsection{MLE-HL}
MLE is a statistical method used to estimate the parameters of a probabilistic model based on observed data. The core idea of MLE is to find the parameter values that maximize the likelihood of the observed data under the estimate. For the task of HL, the likelihood function can be defined as follows:
\begin{equation}
	\mathcal L(\bm{\hat{\theta}})=\sum_i \bigg|\bra{y_i}Me^{-iH(\bm{\hat{\theta}})t}U\ket{0}^{\otimes n}\bigg|^2.
\end{equation}
The summation is taken over all the sampled bit-strings obtained from the measurements. In simulation, the likelihood function is directly optimized using BFGS.

\subsection{Comparison Result}
The results of the benchmark presented in Fig.~\ref{fig:comparison} provide a comprehensive comparison of the iPINN-HL against three baseline methods. Scaling coefficients indicate precision scaling with number of queries. Training times are reported on an NVIDIA RTX 4070 Ti Super GPU. Accuracy metrics (MSE and relative MSE) are computed for $\mathcal{N}(\{x\}) = 10^6$ queries. While other parameters are held the same
as in Sec.~\ref{sec:spinchain} and Appendix~\ref{sec:training_details}.

Across all examined system sizes ($N = 4, 7, 8, 10$) and translation symmetry levels ($s = 1, 2, 4$), iPINN-HL consistently achieves the highest scaling coefficient, approaching the Heisenberg limit ($\mathcal{O}(1/N^2)$). This indicates superior precision scaling with data, where the estimation error decreases quadratically, enabling efficient learning even as the complexity of the quantum system increases. In contrast, DNN-HL and RBM-HL exhibit lower scaling coefficients ranging from $1.501$ to $1.796$, while MLE-HL yields the lowest scaling coefficients ($1.403$ to $1.621$). Notably, iPINN-HL's scaling coefficient remains almost invariant across different $s$, demonstrating robustness to the number of Hamiltonian parameters, whereas the three other baselines show a clear degradation with increasing $s$.

Furthermore, iPINN-HL outperforms the other methods in terms of estimation accuracy, exhibiting the lowest MSE and relative MSE for all configurations considered. For instance, at $N = 7$ and $s = 1$, iPINN-HL achieves an MSE of $1.0 \times 10^{-3}$ and relative MSE of $1.8 \times 10^{-3}$, compared to $3.1 \times 10^{-3}$ and $5.5 \times 10^{-3}$ for DNN-HL, $9.0 \times 10^{-3}$ and $1.6 \times 10^{-2}$ for RBM-HL, and $1.5 \times 10^{-2}$ and $2.7 \times 10^{-2}$ for MLE-HL. This advantage stems from the integration of Schr\"odinger equation into the loss function, which constrains the solution space to physically consistent outcomes and enhances data efficiency. The baselines, lacking such physical priors, exhibit higher errors that increase more rapidly with $N$ and $s$.

While iPINN-HL demonstrates clear superiority in scaling and accuracy, it incurs higher training times compared to DNN-HL and MLE-HL, reflecting the additional computational overhead of incorporating physical constraints. Specifically, the training time for iPINN-HL is elevated relative to DNN-HL due to the evaluation and minimization of the physics loss over collocation points, which involves computing time derivatives of the quantum state and enforcing consistency with the Schr\"odinger equation at multiple sampled times. This adds complexity to each optimization iteration, resulting in times such as $3.3$ hrs for $N = 7, s = 1$, versus $2.8$ hrs for DNN-HL. Similarly, RBM-HL requires longer training owing to its sampling-based nature, where contrastive divergence or Gibbs sampling over the Hilbert space introduces stochastic overhead, leading to times like $5.6$ hrs for $N = 7, s = 1$. ML methods (iPINN-HL, DNN-HL, RBM-HL) generally exceed the non-ML MLE-HL in training duration because they involve differentiation over network parameters in addition to the unknown Hamiltonian parameters, necessitating backpropagation through large models. For MLE-HL, the training time is dominated by repeated matrix exponentials, but BFGS optimization keeps it low because the differentiation only acts on the unknown Hamiltonian parameters without a neural network in the loop, e.g., $1.7$ hrs for $N = 7, s = 4$.

Nevertheless, the generalization capabilities of ML methods, particularly iPINN-HL, mitigate the high cost of data collection in the quantum regime. Traditional approaches like MLE-HL require dense sampling of the dynamics to achieve comparable accuracy, often demanding extensive experimental resources. In contrast, iPINN-HL leverages learned representations and physical priors to extrapolate beyond observed data, reducing the reliance on large datasets and enabling efficient characterization of quantum systems where measurements are expensive or limited. This trade-off, higher computational time for lower data requirements, positions iPINN-HL as a practical tool for advancing quantum technologies, where resource efficiency is paramount.

\begin{figure}
	\includegraphics[width=\linewidth]{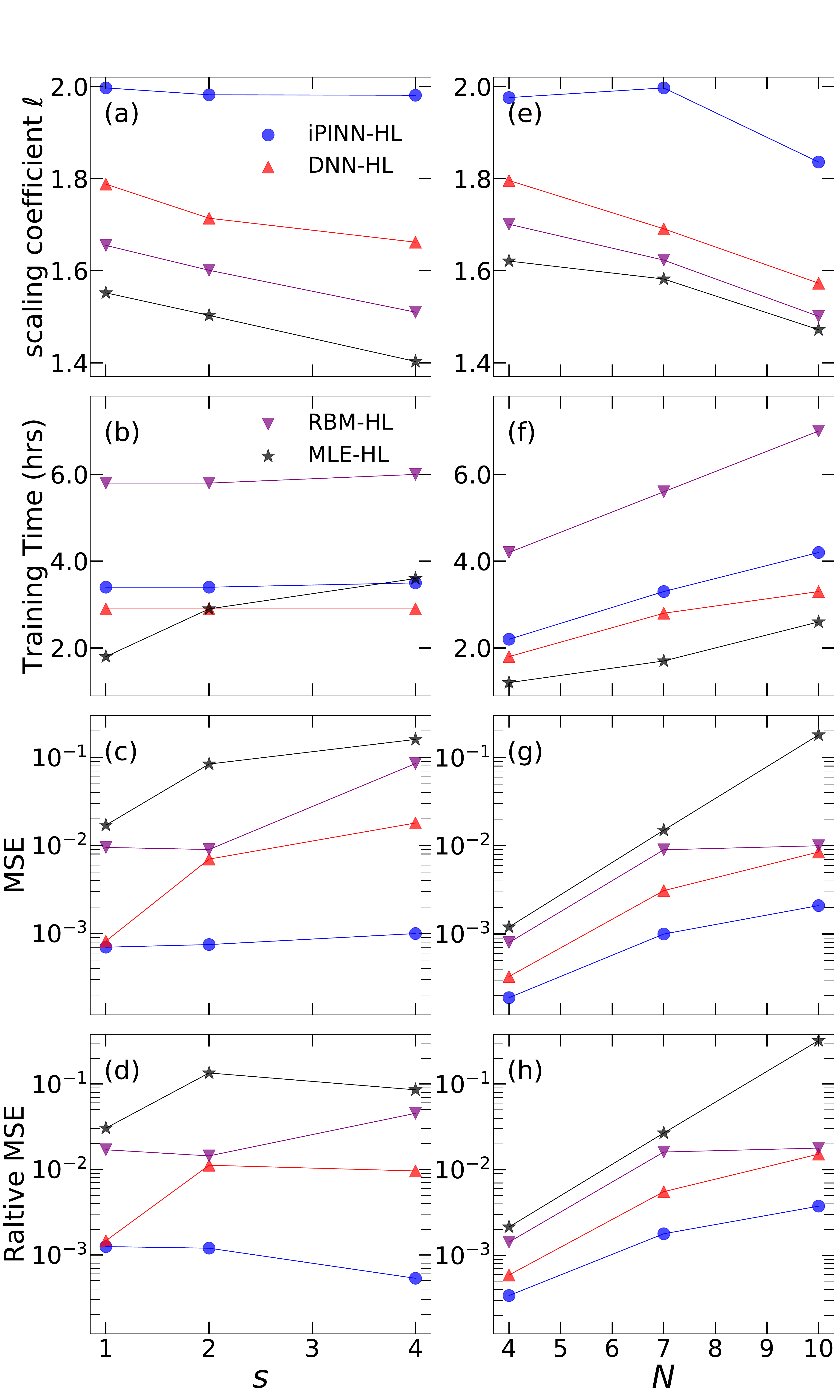}
	\caption{Performance and computational cost of iPINN-HL, DNN-HL, RBM-HL, and MLE-HL for the one-dimensional spin chain model. The left column (panels (a)-(d)) share the same x-axis, namely the translational symmetry factor ($s = 1, 2, 4$) while $N = 8$; the right column (panels (e)-(h))  share the same x-axis, i.e. the number of spins ($N = 4, 7, 10$) while $s = 1$. Each row shares the same y-axis, showing (a,e) the scaling coefficient, (b,f) training hours, (c,g) MSE, and (d,h) relative MSE.}
	\label{fig:comparison}
\end{figure}
\section{Derivation of Noise Channels from the Lindblad Master Equation} \label{appendix.b}

In this appendix, we show the derivations that justify the approximation of solving Lindblad master equation by applying noise channels after unitary gates, in noisy quantum circuit simulations. The goal is to show how continuous-time dynamics can be approximated as discrete quantum channels in a circuit model.

The Lindblad master equation governs the evolution of a quantum system's density matrix $\rho(t)$ under Markovian noise:

\begin{equation}
\frac{d\rho(t)}{dt} = -\frac{i}{\hbar} [H, \rho(t)] + \sum_k \mathcal{D}[L_k]\rho(t),
\end{equation}
where $H$ is the system Hamiltonian, $[H, \rho] = H\rho - \rho H$, and the dissipative term is:
\begin{equation}
\mathcal{D}[L_k]\rho = L_k \rho L_k^\dagger - \frac{1}{2} \{L_k^\dagger L_k, \rho\},
\end{equation}
with $\{A, B\} = AB + BA$ and $L_k$ as the Lindblad operators describing noise processes.

We consider the evolution over a small time step $\Delta t$. The solution to the master equation is:
\begin{equation}
\rho(t + \Delta t) = e^{\mathcal{L}\Delta t} \rho(t),
\end{equation}
where the Liouvillian superoperator is:
\begin{equation}
\mathcal{L}\rho = -\frac{i}{\hbar} [H, \rho] + \sum_k \left( L_k \rho L_k^\dagger - \frac{1}{2} \{L_k^\dagger L_k, \rho\} \right).
\end{equation}

For small $\Delta t$, we approximate:
\begin{equation}
e^{\mathcal{L}\Delta t} \approx 1 + \mathcal{L}\Delta t,
\end{equation}
yielding:
\begin{equation}
\rho(t + \Delta t) \approx \rho(t) + \mathcal{L}\rho(t) \Delta t.
\end{equation}

Now, we split the Liouvillian into coherent and dissipative parts:
\begin{itemize}
    \item Coherent: $\mathcal{L}_{\text{coh}}\rho = -\frac{i}{\hbar} [H, \rho]$,
    \item Dissipative: $\mathcal{L}_{\text{dis}}\rho = \sum_k \left( L_k \rho L_k^\dagger - \frac{1}{2} \{L_k^\dagger L_k, \rho\} \right) $.
\end{itemize}

Using the Trotter-Suzuki approximation for small $\Delta t$:
\begin{equation}
e^{\mathcal{L}\Delta t} \approx e^{\mathcal{L}_{\text{coh}}\Delta t} e^{\mathcal{L}_{\text{dis}}\Delta t} + O(\Delta t^2),
\end{equation}
the coherent part gives unitary evolution:
\begin{equation}
e^{\mathcal{L}_{\text{coh}}\Delta t} \rho = e^{-iH\Delta t/\hbar} \rho e^{iH\Delta t/\hbar}.
\end{equation}
Since typical quantum gate durations $t_g$ are much shorter than the characteristic coherence times of the system (for superconducting qubits, $t_g \sim 10\text{--}100\,\mathrm{ns}$ while $T_{1,2} \sim 10\text{--}100\,\mu\mathrm{s}$), the probability of a decoherence event within a single gate is very small. This justifies treating the gate duration as a short-time step in the Trotter expansion. Thus, we identify $\Delta t = t_g$, so that the coherent unitary dynamics and the dissipative noise can be separated to first order. 
For a gate time $t_g = \Delta t$, the unitary gate is $U = e^{-iHt_g/\hbar}$, so:
\begin{equation}
\rho \to U \rho U^\dagger.
\end{equation}

The dissipative part is approximated as:
\begin{equation}
e^{\mathcal{L}_{\text{dis}}t_g} \rho \approx \rho + t_g \sum_k \left( L_k \rho L_k^\dagger - \frac{1}{2} \{L_k^\dagger L_k, \rho\} \right).
\end{equation}

This can be written as a quantum channel:
\begin{equation}
\rho \to \mathcal{E}(\rho) = \sum_k K_k \rho K_k^\dagger,
\end{equation}
with Kraus operators:
\begin{itemize}
    \item $K_0 = I - \frac{t_g}{2} \sum_k L_k^\dagger L_k$,
    \item $K_k = \sqrt{t_g} L_k$ for each $L_k$.
\end{itemize}
The Kraus operators satisfy $\sum_k K_k^\dagger K_k = I$ to first order, ensuring a trace-preserving map.

In a quantum circuit, we approximate the evolution over a gate time $t_g$ as a unitary gate followed by a noise channel:
\begin{equation}
\rho \to \mathcal{E}(U \rho U^\dagger),
\end{equation}
where $U = e^{-iHt_g/\hbar}$ and $\mathcal{E}$ is the noise channel derived from the dissipative terms over $t_g$.

As an example, we consider a single qubit with depolarizing noise. The master equation includes:
\begin{equation}
\frac{d\rho}{dt} = -\frac{i}{\hbar} [H, \rho] + \frac{\gamma}{4} \sum_{i=1}^3 \left( \sigma_i \rho \sigma_i - \frac{1}{2} \{\sigma_i^2, \rho\} \right),
\end{equation}
where $\sigma_i = \{X, Y, Z\}$ and $\sigma_i^2 = I$. The dissipative term becomes:
\begin{equation}
\mathcal{D}[\sigma_i]\rho = \sigma_i \rho \sigma_i - \rho.
\end{equation}

Over gate time $t_g$, the noise channel is:
\begin{equation}
\mathcal{E}_{\text{dis}}(\rho) = (1 - p) \rho + \frac{p}{3} \sum_{i=1}^3 \sigma_i \rho \sigma_i,
\end{equation}
with $p = 3\gamma t_g / 4$. The Kraus operators are:
\begin{itemize}
    \item $K_0 = \sqrt{1 - p} I$,
    \item $K_i = \sqrt{p/3} \sigma_i$ for $i = 1, 2, 3$.
\end{itemize}

The full evolution after a gate is:
\begin{equation}
\rho \to (1 - p) U \rho U^\dagger + \frac{p}{3} \sum_{i=1}^3 \sigma_i U \rho U^\dagger \sigma_i.
\end{equation}

This isotropic depolarizing channel is equivalent to the heuristic model where, with probability $p$, the state is replaced by the maximally mixed state $I/d$. Indeed, the map can be written as $\mathcal{E}(\rho) = (1-p)\rho + p \, I/d$, which matches the action of the depolarizing channel derived from the Lindblad form. This matches the model used in Sec.~\ref{sec:CR_gate}. However, for non-Markovian noise or long gate times, the full master equation must be solved numerically.

\section{Training Details of iPINN-HL and DNN-HL} \label{sec:training_details}
In TABLE~\ref{tab:training_details} we present the training details of both iPINN-HL and DNN-HL presented in the main text.
\begin{table*}[ht!]
\centering
\caption{Training details for iPINN-HL and DNN-HL. Training will be early stopped after 200 epochs of no validation loss improvement. Here, lr (learning rate) sets the step size of updates, $\beta_1$ (first moment decay) controls the smoothing of momentum, $\beta_2$ (second moment decay) regulates the smoothing of variance, and $\epsilon$ (stability constant) prevents division by zero.}
\label{tab:training_details}
\begin{tabular}{lcc}
\hline
Method & iPINN-HL & DNN-HL \\
\hline
Network Structure & Feedforward NN, 5 layers (120 nodes each) & Feedforward NN, 5 layers (120 nodes each)\\
Activation Function & SiLU \cite{hendrycks2016gaussian} & Tanh \\
Optimizer & ADAM (lr=0.001, $\beta_1=0.9$, $\beta_2=0.999$,   & ADAM (lr=0.001, $\beta_1=0.9$, $\beta_2=0.999$, \\
&$\epsilon=10^{-8}$) + BFGS& $\epsilon=10^{-8}$)\\
Training Epochs & 3000--5000, early stopping & 3000--5000, early stopping \\
\hline
\end{tabular}
\end{table*}
} 

\end{document}